\documentclass[a4paper,twocolumn,aps,prb]{revtex4-1}

\newcommand{\eq}[1]{\begin{equation}#1\end{equation}}
\newcommand{\dd}{\mathrm{d}}
\newcommand{\ee}{\mathrm{e}}

\newcommand{\Tr}{\mathrm{Tr \,}}
\newcommand{\Trb}{\mathrm{Tr}_B \,}

\newcommand{\ptr}{\rho^{T_2}_{A}}
\newcommand{\lneg}{\mathcal{E}}
\newcommand{\twn}{\mathcal{T}_n}
\newcommand{\twnb}{\overline{\mathcal{T}}_n}
\newcommand{\FF}{\mathrm{\mathcal{F}}}
\newcommand{\TT}{\mathrm{\mathcal{T}}}
\newcommand{\ddn}{\Delta^{(2)}_n}

\newcommand{\bea}{\begin{eqnarray}}
\newcommand{\eea}{\end{eqnarray}}
\newcommand{\beas}{\begin{eqnarray*}}
\newcommand{\eeas}{\end{eqnarray*}}

\newcommand{\twomat}[4]{\left(\begin{array}{cc} #1 & #2 \\ #3 & #4\end{array}\right)}
\newcommand{\identity}{\openone}

\usepackage{amsmath}
\usepackage{amsfonts}
\usepackage{graphics}
\usepackage{graphicx}
\usepackage{psfrag}
\usepackage{color}
\usepackage{colordvi}
\usepackage{bbm}

\begin{document}

\title{Entanglement negativity in two-dimensional free lattice models}

\author{Viktor Eisler$^{1,2}$ and Zolt\'an Zimbor\'as$^{3}$}
\affiliation{
$^1$Institut f\"ur Theoretische Physik, Technische Universit\"at Graz, Petersgasse 16, A-8010 Graz, Austria \\
$^2$MTA-ELTE Theoretical Physics Research Group,
E\"otv\"os Lor\'and University, P\'azm\'any P\'eter s\'et\'any 1/a, H-1117 Budapest, Hungary \\
$^3$Department of Computer Science, University College London, Gower Street,
{WC1E 6BT} London, United Kingdom}

\begin{abstract}
We study the scaling properties of the ground-state entanglement between finite subsystems of infinite two-dimensional free lattice models,
as measured by the logarithmic negativity. For adjacent regions with a common boundary, we observe that the negativity follows a strict
area law for a lattice of harmonic oscillators, whereas for fermionic hopping models the numerical results indicate a multiplicative logarithmic
correction. In this latter case, we conjecture a formula for the prefactor of the area-law violating term, which is entirely determined by the
geometries of the Fermi surface and the boundary between the subsystems. The conjecture is tested against numerical results
and a good agreement is found.\end{abstract}

\maketitle

\section{Introduction}

In recent years, ideas from quantum information theory have stimulated major developments in the field of strongly correlated systems. The entanglement properties of many-body states lies at the center of these studies. An important insight in this context is that, for ground states of local Hamiltonians, the entanglement between a subsystem and the rest of the system obeys an area law with a possible multiplicative logarithmic correction.\cite{ECP09,CCD09} Moreover, the details of the ground-state entanglement scaling carries important information about the system, e.g., one can determine the universality class of one-dimensional critical models \cite{Vidal03, CC04} or detect topological order.\cite{Kitaev06,LevinWen06, Hamma05}

While there has been numerous studies on the entanglement between a subsystem and its complement, much less is known about the entanglement between two regions that together do not constitute the entire system. The main reason is that the von Neumann entropy cannot be applied to study this case, since it is a measure of entanglement only for bipartite pure states. In the case of non-complementary regions, embedded in a larger system, one needs a different characterization, because the state reduced to the union of the subsystems is in general mixed. Among the various entanglement monotones for mixed states, entanglement negativity turns out to be a particularly useful measure.\cite{Kim00,VW02,Plenio05} It is easily computable for bosonic Gaussian states, \cite{Simon00, AEPW02} and recently also some results concerning fermionic Gaussian states have appeared.\cite{EZ15} Using these Gaussian methods, the tripartite ground-state entanglement negativity has been recently investigated for one-dimensional free bosonic\cite{CCT12,CCT13,NCT15} and fermionic models.\cite{EZ15,CTC15,CTC15b}

In this paper, we continue these surveys by considering two-dimensional lattices. When comparing the entanglement content of bosonic systems with that of fermionic ones, the dimensionality plays an important role.  For critical one-dimensional systems, the entanglement between an interval and the rest of the chain scales logarithmically with the length of the interval both for fermions and bosons. However, in higher dimensions, harmonic lattices (which can be viewed as free boson models) obey a strict area law, \cite{PEDC05, CEP07} while free fermions may violate the area law by a multiplicative logarithmic correction. \cite{Wolf06,GK06,BCS06,FZ07} There is an appealing physical picture that gives an intuitive understanding of  this difference. \cite{Swingle10} For fermion models, the occupied and unoccupied modes in momentum space are separated by the Fermi-surface, characterized by a vanishing excitation gap. In the generic case, this is a $(d {-}1)$-dimensional surface of the $d$-dimensional Brillouin zone, around which the dispersion can be linearized. One can think that each patch of the Fermi surface is equivalent to a single gapless excitation, described by a  $1 {+} 1$ dimensional conformal field theory (CFT), which then leads to a multiplicative logarithmic correction to the area law. Indeed, the above argument implies exactly an entanglement entropy scaling that was already obtained based on the Widom conjecture.\cite{GK06} In contrast, for harmonic lattices, the gapless bosonic modes are supported only on a single point (or on a discrete number of points) in the momentum space, which can give at most an additive logarithmic contribution to the entanglement entropy.

Here we set out to investigate the validity of the above intuitive physical picture also for the logarithmic negativity. We find that again a strict area law holds for harmonic oscillator systems, while in the case of free fermions our results indicate a multiplicative logarithmic correction. Moreover, using results obtained within a 1+1 dimensional CFT framework as an input,\cite{CCT12,CCT13} we formulate a simple conjecture for the area-law violating term of the entanglement negativity, and present numerical evidence in its favor.

The paper is structured as follows. In Section~\ref{sec:neg} we recall the definition of logarithmic negativity and discuss how the partial transposition operation acts on bosonic and fermionic Gaussian states. CFT techniques concerning negativity are shortly reviewed in Section~\ref{sec:CFT}, which are then used to calculate the negativity scaling for different subsystem geometries. The main results of the paper on two-dimensional models are presented in Section~\ref{sec:2d}. We conclude in Section~\ref{sec:Discussion} with a short discussion of the results and their possible extensions. Various details of the analytical calculations are included in the two Appendices.

\section{Partial transpose and logarithmic negativity}\label{sec:neg}

\begin{figure}[thb]
\center
\includegraphics[width=\columnwidth]{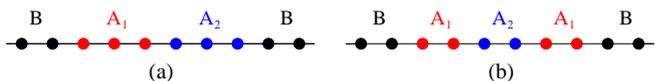}
\caption{Different tripartitions of a 1D chain.}
\label{fig:1d}
\end{figure}

We will consider the ground state $\rho$ of a many-body system defined on a lattice
which is subdivided into three disjoint subsets $A_1$, $A_2$ and $B$, such that
$A=A_1 \cup A_2$ and $A \cup B$ corresponds to the entire lattice. In case of a
one-dimensional chain, two such tripartitionings are illustrated on Fig.~\ref{fig:1d}.
The reduced density matrix (RDM) of subsystem $A$ is given by $\rho_A = \Trb  \rho$
and its partial transpose w.r.t $A_2$ is defined through its matrix elements as
\eq{
\langle e_i^{(1)} e_j^{(2)} | \rho_A^{T_2} | e_k^{(1)} e_l^{(2)} \rangle =
\langle e_i^{(1)} e_l^{(2)} | \rho_A | e_k^{(1)} e_j^{(2)} \rangle,
\label{pt}
}
where $\{ | e_i^{(1)}\rangle \}$ and $\{ | e_j^{(2)}\rangle \}$ denote complete bases 
spanning the Hilbert spaces pertaining to subsets $A_1$ and $A_2$. 

%
%

In general, the result of the partial transposition is not a positive operator and the appearance of negative eigenvalues signals entanglement.\cite{Peres96,H3} Based on this property, a suitable entanglement measure called logarithmic negativity \cite{VW02} can be introduced as
\eq{
\lneg = \ln \Tr |\ptr| .
\label{ln}}

Despite being known as a computable measure of entanglement, the logarithmic negativity requires
the knowledge of the full spectrum of the partial transpose $\ptr$, which, in practice, is difficult
to obtain for large many-body systems. A well-known exception is the class of Gaussian states.
In this section we will review Gaussian techniques, which will be used later in Section~\ref{sec:2d}.

\subsection{Bosonic Gaussian states}

Considering systems that are  defined by
a set of canonical coordinates $\{x_n\}$ and momenta $\{p_n\}$ with $n=1,\dots,N$ indexing
the modes (or lattice sites), one can define 
continuous variable Gaussian states, also known as
bosonic Gaussian states. Introducing the notation $R_{2n-1}=x_n$ and $R_{2n}=p_n$, bosonic Gaussian 
states are uniquely defined via their covariance matrix $\Gamma_{kl}=\langle \{ R_k, R_l \} \rangle$,
with higher order correlation functions factorizing according to Wick's theorem.

The reduced density matrix $\rho_A$ of a Gaussian state is again Gaussian and characterized by
the reduced covariance matrix $\Gamma_A$, where the indices are restricted to the subset $A$.
Furthermore, the partial transposition has a particularly
simple action on these states, since it can be represented as a partial time-reversal, flipping the sign
of the momenta in the corresponding subsystem while leaving the coordinate variables unchanged. \cite{Simon00}
In turn, the partial transpose $\ptr$ of the RDM is a Gaussian operator with covariance matrix \cite{AEPW02}
\eq{
\Gamma^{T_2}_A = R_{A_2} \Gamma_A R_{A_2},
\label{covb}
}
where
\eq{
R_{A_2} =\bigoplus_{A_1}\twomat{1}{}{}{1}\bigoplus_{A_2}\twomat{1}{}{}{-1}
}
is the diagonal matrix reversing the momenta in $A_2$.

Due to its Gaussianity, one has direct access to the full spectrum of the partial transpose
via the symplectic spectrum $\{\nu_j\}$ of $\Gamma^{T_2}_A$. In particular, the formula
for the logarithmic negativity in Eq.~\eqref{ln} can directly be evaluated as
\eq{
\mathcal{E} = - \sum_{j=1}^{|A|} \ln \min (\nu_j,1).
\label{lnbg}}
This formula has been used in the earlier studies of negativity in various Gaussian many-body
states.\cite{FCGA08,AW08,Anders08,MRPR09}

\subsection{Fermionic Gaussian states\label{fgs}}

Similarly to the bosonic case, the fermionic version of Gaussian states can
also be defined, pertaining to a lattice system with creation $c_n^\dag$ and annihilation
operators $c_n$ satisfying canonical anticommutation relations $\{c_m^\dag,c_n\}=\delta_{m,n}$.
For a completely analogous treatment with the bosonic case, one can introduce Majorana
operators $a_{2n-1}=c_n+c_n^\dag$ and $a_{2n}=i(c_n-c_n^\dag)$ and define the fermionic
covariance matrix as $\Gamma_{kl}=\langle \left[ a_k, a_l \right] \rangle /2$.
These two-point functions completely characterize a fermionic Gaussian state, as the 
higher-order correlations are given by the fermionic version of Wick's theorem.

Identically to its bosonic counterpart, the reduction of a fermionic Gaussian state to a subsystem $A$ remains Gaussian with reduced covariance matrix $\Gamma_A$. In sharp contrast, however, the partial transpose operation for fermions does not preserve Gaussianity. Nonetheless, it has been shown in Ref.~[\onlinecite{EZ15}], that in a suitable basis the partial transpose of a Gaussian RDM can be decomposed as the linear combination of two Gaussian operators. 
Indeed, the partial transposition with respect to $A_2$ leaves the modes in $A_1$ invariant and acts only on the ones in $A_2$.
Considering a product of $n$ distinct Majorana operators $M=a_{i_1} a_{i_2} \cdots a_{i_n}$ from subsystem $A_2$,
the transposition, in a particular basis, acts as $M^{T_2} = ({-}1)^{f(n)}M$ where
\eq{
f(n)=
\begin{cases}
0 & \! \! \! \text{if} \; n \! \! \! \! \! \mod 4 \in \{0,1\},\\
1 & \! \! \! \text{if}  \; n \! \! \! \! \!\mod 4 \in \{2,3\}.\\
\end{cases} 
}
Using this definition, the partial transpose of a Gaussian RDM can be written in the form \cite{EZ15}
\eq{
\ptr= \frac{1-i}{2} O^+_A + \frac{1+i}{2} O^-_A.
\label{ptrho}
}
Here $O^{\pm}_A$ are Gaussian operators  with covariance matrices $\Gamma^{\pm}_A$
that are defined as
\eq{
\Gamma^{\pm}_A = T^{\pm}_{A_2} \Gamma_A T^{\pm}_{A_2},
\label{covf}}
where
\eq{
T^{\pm}_{A_2} =\bigoplus_{A_1}\twomat{1}{}{}{1}\bigoplus_{A_2}\twomat{\pm i}{}{}{\pm i}.
}

Although the spectra of $O^{\pm}_A$ can be constructed explicitly, the two operators do not
commute in general and one has no direct access to the eigenvalues of $\ptr$ and,
as a consequence, to the logarithmic negativity. Nevertheless, one can still extract some
useful information from this form of the partial transposed RDM: the traces of its moments,
i.e. $\Tr (\ptr)^n$. Indeed, factoring out Eq.~\eqref{ptrho}, one is left with a sum of traces of
products of Gaussian operators, each of which can be calculated explicitly.
The steps of this procedure are summarized in Appendix~\ref{sec:appa}.

The moments of the partial transpose, despite not being entanglement measures,
are the basic objects that are also attainable in CFT and were successfully used to characterize
tripartite entanglement in 1D critical systems. \cite{CCT12,CCT13, EZ15}
In this paper, we will also study quantities related to the moments to obtain an indication
of the negativity scaling for 2D free fermion systems. To understand the role of the subsystem
geometry in the 2D case, we first give a brief overview of the method employed in CFT to extract
the moments of the partial transpose, and then apply it to simple 1D subsystem arrangements.

\section{Entanglement negativity in CFT}\label{sec:CFT}

Conformal field theory provides a powerful machinery for the unraveling of universal properties
of negativity scaling. Using CFT techniques, one could investigate the negativity in critical ground states,
\cite{CCT12, CCT13, NCT15,CTC15,CTC15b} in low temperature Gibbs states,\cite{EZ14,CCT15}
and even in non-equilibrium situations.\cite{EZ14, CTC14, HD15, WCR15} 
Moreover, some recent progress has been made in extending the technique to massive quantum field theories. \cite{BCD15}
In this section we shortly review the main tools needed for such calculations,
and use them to compare the negativity scaling for two different subsystem geometries depicted in Fig.~\ref{fig:1d}.
These results also constitute an essential input for our later studies of 2D free fermion models in Section~\ref{sec:2d}.

\subsection{The replica trick}

The calculation of entanglement negativity in CFT relies on the path-integral
representation of the partial transpose and on a clever application of the replica trick. \cite{CCT12,CCT13}
In the first step, one defines the ratio of the moments of the RDM and its partial transpose,
as well as its logarithm
\eq{
R_n = \frac{\Tr (\ptr)^n}{\Tr \rho^n_A}, \qquad
\lneg_n = \ln R_{n}.
\label{rn}}
Now, the crucial observation is that the ratio $R_n$ has a strong parity dependence
due to the presence of negative eigenvalues of $\ptr$. In particular, the trace norm
in Eq.~\eqref{ln} can be recovered by considering the series $\lneg_{n_e}$ on even integers
$n_e$ and taking the limit
\eq{
\lneg = \lim_{n_e \to 1} \lneg_{n_e}.
\label{lnlim}}

We note that, instead of the ratio $R_n$, one could simply use the moments of the partial transpose
and their logarithms to obtain the entanglement negativity from the same limit as in \eqref{lnlim}.
However, the definition of $R_n$ turns out to be more useful in various situations. On one hand,
these ratios were shown to be universal in case of two non-adjacent intervals, i.e., $R_n$ depends
only on the four-point ratio of the intervals through a universal scaling function corresponding to
the given CFT.\cite{CCT13, Alba13}
On the other hand, while the same is not true for adjacent intervals, the ratios $R_n$ have a much
clearer interpretation also in this case, as will be shown below.

In order to carry out the limit \eqref{lnlim}, one needs an explicit formula for $\lneg_{n}$
and thus a method to calculate the traces in Eq.~\eqref{rn}. This can be done by rewriting them
as expectation values of products of twist fields $\twn$ and $\twnb$, permuting cyclically or
anti-cyclically between the replicas. When both $A_i=\left[u_i,v_i\right]$ with $i=1,2$ correspond to a
single interval, one has \cite{CCT13}
\eq{
\begin{split}
\Tr \rho^n_A = \langle \twn(u_1) \twnb(v_1) \twn(u_2) \twnb(v_2) \rangle, \\
\Tr (\ptr)^n = \langle \twn(u_1) \twnb(v_1) \twnb(u_2) \twn(v_2) \rangle.
\end{split}
}
In other words, when considering moments of the RDM, the twist fields $\twn$ and $\twnb$
have to be inserted at the start- and endpoints of the slits corresponding to $A_1$ and $A_2$.
For the partial transpose, the edges of the slit $A_2$, and thus the corresponding twist field insertions
have to be interchanged. 
Analogously, considering $N$ non-intersecting intervals $[u_i, v_i]$ for $i=1, \ldots N$,
with $u_i< v_i$ and $v_i \le u_{i+1}$, we can split them into two complementary sets
$I_1$ and $I_2 = \{1, \ldots , N\} \setminus I_1$ which define the subsystems
$A_j = \cup_{i \in I_j} [u_i, v_i]$ for $j=1,2$. The moments are then given by
\begin{equation}
\begin{split}
\Tr \rho^n_A &= \langle \prod_{i=1}^{N}\twn(u_i) \twnb(v_i) \rangle, \\
\Tr (\ptr)^n &= \langle \prod_{i=1}^{N} 
\mathcal{S}_2\left[\twn(u_i) \twnb(v_i)\right] \rangle,
\end{split} \label{eq:manyintervals}
\end{equation}
where the partial swap operator $\mathcal{S}_2$ acts as
\begin{equation}
\mathcal{S}_2\left[\twn(u_i) \twnb(v_i)\right] = 
\begin{cases}
 \twn(u_i) \twnb(v_i) \; \; \; \text{if} \; \; i \in I_1, \\
\twnb(u_i) \twn(v_i) \; \; \; \text{if} \; \; i \in I_2.
\end{cases}
\end{equation}
It should be mentioned that the general structure of these $2N$-point functions of twist fields
becomes rather involved for $N>2$ and analytical results are only available for some special CFTs. \cite{CTT14}

\subsection{Adjacent intervals}

We first consider the simplest situation with two adjacent intervals of lengths $\ell_1$ and $\ell_2$, within an infinite one-dimensional critical system.\cite{CCT13}
One can then set $u_1=-\ell_1$, $v_2=\ell_2$ and $v_1=u_2=0$, hence $R_n$ can be written as
\eq{
R_n = \frac{\langle \twn(-\ell_1) \twnb^2(0) \twn(\ell_2)\rangle}{\langle \twn(-\ell_1) \twnb(\ell_2)\rangle},
\label{rnai}
}
and thus as the ratio of a three-point and a two-point function on the full complex plane.
It is well known, that the twist fields $\twn$ and $\twnb$ behave like primary operators with
scaling dimension \cite{CC04}
\begin{equation}
\Delta_n = \frac{c}{12}\left( n - \frac{1}{n} \right).
\end{equation}
In contrast, the numerator of Eq.~\eqref{rnai} contains an insertion of a squared twist field
$\twnb^2$, whose scaling dimension shows a strong dependence on the number of replicas \cite{CCT13}
\begin{equation}
\ddn=
\begin{cases}
\Delta_{n_o} & n=n_o, \\
2\Delta_{n_e/2} & n=n_e.
\end{cases}
\end{equation}
Indeed, in case of $n=n_e$ even, the actions of $\twn^2$ and $\twnb^2$ completely decouple the
even and odd layers of replica sheets.

Finally, one can use the CFT results for the two-point function
\eq{
\langle \twn(-\ell_1) \twnb(\ell_2)\rangle = c_n (\ell_1+\ell_2)^{-2\Delta_n} \, ,
}
where $c_n$ are non-universal constants. Similarly, the three-point function
follows from conformal symmetry as
\begin{align}
&\langle \twn(-\ell_1) \twnb^2(0) \twn(\ell_2)\rangle = \nonumber \\
&c^2_n C_{\twn\twnb^2\twn} (\ell_1+\ell_2)^{-2\Delta_n}\left(\frac{\ell_1\ell_2}{\ell_1+\ell_2}\right)^{-\ddn},
\end{align}
where $C_{\twn\twnb^2\twn}$ are universal structure constants.
Substituting into \eqref{rnai} and taking the logarithm, one arrives at
\eq{
\lneg_{n}= -\ddn \ln \frac{\ell_1 \ell_2}{\ell_1+\ell_2} + const. \, ,
\label{en1dadj}
}
which manifestly depends only on the scaling dimension $\ddn$.
The entanglement negativity follows through the limit \eqref{lnlim} as
\eq{
\lneg=\frac{c}{4} \ln \frac{\ell_1 \ell_2}{\ell_1+\ell_2} + const.
\label{ln1dadj}
}

\subsection{Embedded geometry}

In the case of an embedded geometry depicted in Fig.~\ref{fig:1d}(b), one can perform 
a similar analysis as before. Choosing the subsystems as $A_1=[u_1, v_1] \cup [u_2, v_2]$
and $A_2=[v_1, u_2]$, the moments of the RDM and its partial transpose can be read
off from Eq.~\eqref{eq:manyintervals}
\begin{align}
R_n 
= \frac{\langle \twn(u_1) \twnb^2(v_1)  \twn^2(u_2) \twnb(v_2)\rangle}{\langle \twn(u_1) \twnb(v_2)\rangle}. \label{eq:embed_ratio}
\end{align}
Due to global conformal invariance, the four point function appearing in \eqref{eq:embed_ratio}
has the form \cite{CCT15,HD15}
\begin{align} 
&\langle  \TT_n(u_1) \overline{\TT}_n^2(v_1)  \TT^2_n(u_2) \overline{\TT}_n(v_2) \rangle=
\nonumber \\
&c_n (v_2 - u_1)^{-2 \Delta_n}(u_2 - v_1)^{-2\Delta^{(2)}_n} \eta^{-\Delta_n^{(2)}} \FF_n(\eta),  \label{eq:4point_CFT}
\end{align}
where $\FF_n$ is a scaling function  depending only on the four-point ratio $\eta$ defined as
\begin{equation}
\eta= \frac{(v_1 - u_1)(v_2-u_2)}{(u_2-u_1)(v_2-v_1)}.
\end{equation} 
In writing Eq.~\eqref{eq:4point_CFT} we have separated a term that
becomes divergent in the limit $\eta \to 0$, a behavior which can be found
from the operator product expansion (OPE) technique. \cite{CCT15,HD15}
Although the precise form of $\FF_n(\eta)$ is not known, such a definition ensures
that it tends to constant values in both limits $\eta \to 0$ and $\eta \to 1$.

We will consider a symmetric embedding, i.e., $v_2-u_2=v_1-u_1=\ell_1$ and
$u_2-v_1=\ell_2$, hence the four point-ratio is given by
\eq{
\eta = \left(\frac{\ell_1}{\ell_1+\ell_2}\right)^2 \, .
\label{eta}}
Dividing \eqref{eq:4point_CFT} by the two-point function, one obtains for the logarithmic ratio
\eq{
\lneg_n= - 2 \Delta_n^{(2)}\ln \frac{\ell_1 \ell_2}{\ell_1+\ell_2} + \ln \FF_n(\eta),
\label{en1demb}}
which, for $\eta$ fixed and $\ell_1,\ell_2 \gg 1$, diverges logarithmically with a doubled prefactor
compared to the result \eqref{en1dadj} for adjacent intervals. In particular for $\eta \to 0$,
i.e., for large separations $\ell_2 \gg \ell_1$ between the two intervals of $A_1$, one has
$\FF_n(0)=C_{\twn \twnb^2\twnb}^2$ and thus the subleading term depends only on the structure
constants of the corresponding OPE. Therefore, the entanglement negativity in the embedded geometry
will be asymptotically twice as large as for adjacent intervals in Eq.~\eqref{ln1dadj},
due to the two contact points between $A_1$ and $A_2$. This is very reminiscent of
the behavior of bipartite R\'enyi entropies with periodic vs. open boundary conditions. \cite{CC04}

\section{Entanglement negativity in 2D}\label{sec:2d}

In the previous section it was recalled that for critical 1D systems the area law for negativity is violated
logarithmically, and we also derived how the prefactor of this scaling depends on the subsystem geometry.
In this section the analogous questions for 2D systems will be studied.
In the bipartite case, the simple general connection between criticality and ground-state entanglement properties is lost.
In fact, when considering the scaling of entanglement entropy of a subsystem, for many critical 2D systems, such as the
harmonic lattice \cite{BCS06} or the Heisenberg model, \cite{Melko12} a strict area law holds. For other critical models,
e.g., for free fermions\cite{Wolf06, GK06,LSS14} or interacting Fermi liquids,\cite{DSY12, MT13}
multiplicative logarithmic corrections to the area law can still persist.

The above anomaly for fermionic models has its roots in the presence of a Fermi-surface, and its precursor can be traced back to 1D systems. Indeed, if the ground state of a free fermion chain is given by several disconnected Fermi seas instead of a single one, then the entanglement entropy between an interval of length $\ell$ and the rest of the chain gets multiplied by the number of Fermi points (boundaries of the Fermi seas).\cite{KM05,EZ05,KZ10,AEFS14} Therefore, each Fermi point and subsystem boundary lends a $1/12 \ln \ell$ contribution of an independent chiral CFT to the entanglement. Hence the overall entropy is proportional both to the number of momentum-space boundaries of the Fermi seas and the real-space boundaries of the subsystem. 

This argument can now be lifted to $d$-dimensional systems as follows.\cite{Swingle10, Swingle12,Sorkin15} Consider a ground state defined by the occupied fermionic modes whose border is given by a $(d{-}1)$-dimensional Fermi surface in momentum space, and suppose we are interested in the bipartite entanglement of a spatial region of linear extent $\ell$.
If each patch of the Fermi surface is considered as a source of chiral CFT excitations, with the direction of Fermi velocity
given by the normal vector ${\bf n}_{\bf q}$, then its entangling contribution along spatial direction ${\bf n_{\bf r}}$
should be proportional to $| {\bf n}_{\bf q} {\bf n}_{\bf r} |$. Summing up the contributions from the different
patches of the Fermi surface $\partial F$ and the real-space boundaries $\partial A$, one arrives at
\begin{equation}\label{eq:dD}
S= \frac{1}{12} \ln \ell \cdot \frac{\ell^{d-1}}{(2\pi)^{d-1}} \int_{\partial F} \dd S_{\bf q}
\int_{\partial A} \dd S_{\bf r} |\mathbf{n}_{\bf q} \mathbf{n}_{\bf r}| \, ,
\end{equation} 
where the linear size of region $A$ has been scaled out from the integral and
the proper measure on the Fermi-surface has been taken into account.

Remarkably, the very simple argument leading to Eq.~\eqref{eq:dD} gives the precise asymptotics
of the area-law violating term in the entanglement entropy which has been tested numerically for a
number of (2D or 3D) free-fermion systems \cite{BCS06,LDYRH06,CMV12,RSLHS13} and recently even proved
rigorously.\cite{LSS14} Moreover, it also accounts for the observation that the area law in 2D is restored
whenever the Fermi surface degenerates to a number of points,\cite{BCS06,LDYRH06}
since $\partial F$ is of zero measure in the integral \eqref{eq:dD} of the anomalous term.
In fact, this is the very situation also for a harmonic lattice with short-ranged interactions, where
the gapless bosonic modes are usually supported only on a finite number of points in momentum space.
As a further proof of consistency, one should mention that some exotic Bose liquids, featuring a
bosonic analogue of a Fermi surface, lead again to logarithmic violations of the area law. \cite{LYB13,LY15}

\begin{figure}[thb]
\center
\includegraphics[width=.7\columnwidth]{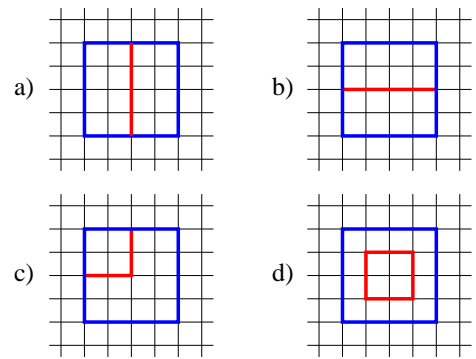}
\caption{Different choices for subsystems $A_1$ and $A_2$ with their common boundary
$\partial A_{12}$ shown in red. The blue rectangle corresponds to the boundary $\partial A$
of subsystem $A=A_1 \cup A_2$.}
\label{fig:2d}
\end{figure}

Returning to the case of entanglement negativity, we expect that the same picture should be valid,
which is supported by the numerics presented in this section. For harmonic lattices we find that a
strict area law holds, while for fermionic systems the results are consistent with a multiplicative
logarithmic correction, given by a formula that extends CFT results in the spirit of Eq.~\eqref{eq:dD}.
To test its validity, we will both consider different subsystem geometries, as shown in Fig.~\ref{fig:2d},
and different Fermi surfaces tuned by the anisotropy of the lattices.

%
%

\subsection{Harmonic lattice}

%
\begin{figure*}[t]
\psfrag{E/L}[][][.9]{$\lneg/\ell$}
\psfrag{L}[][][.9]{$\ell$}
\center
\includegraphics[width=.9\columnwidth]{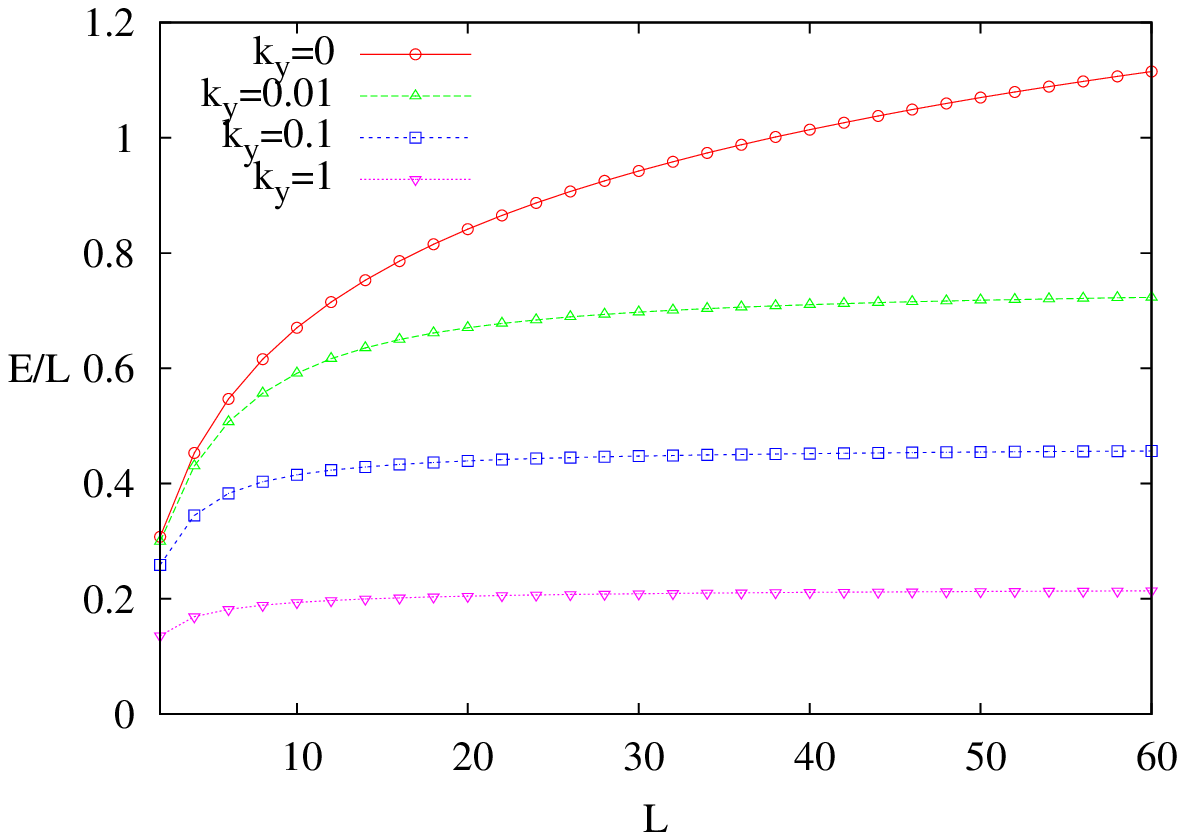}
\includegraphics[width=.9\columnwidth]{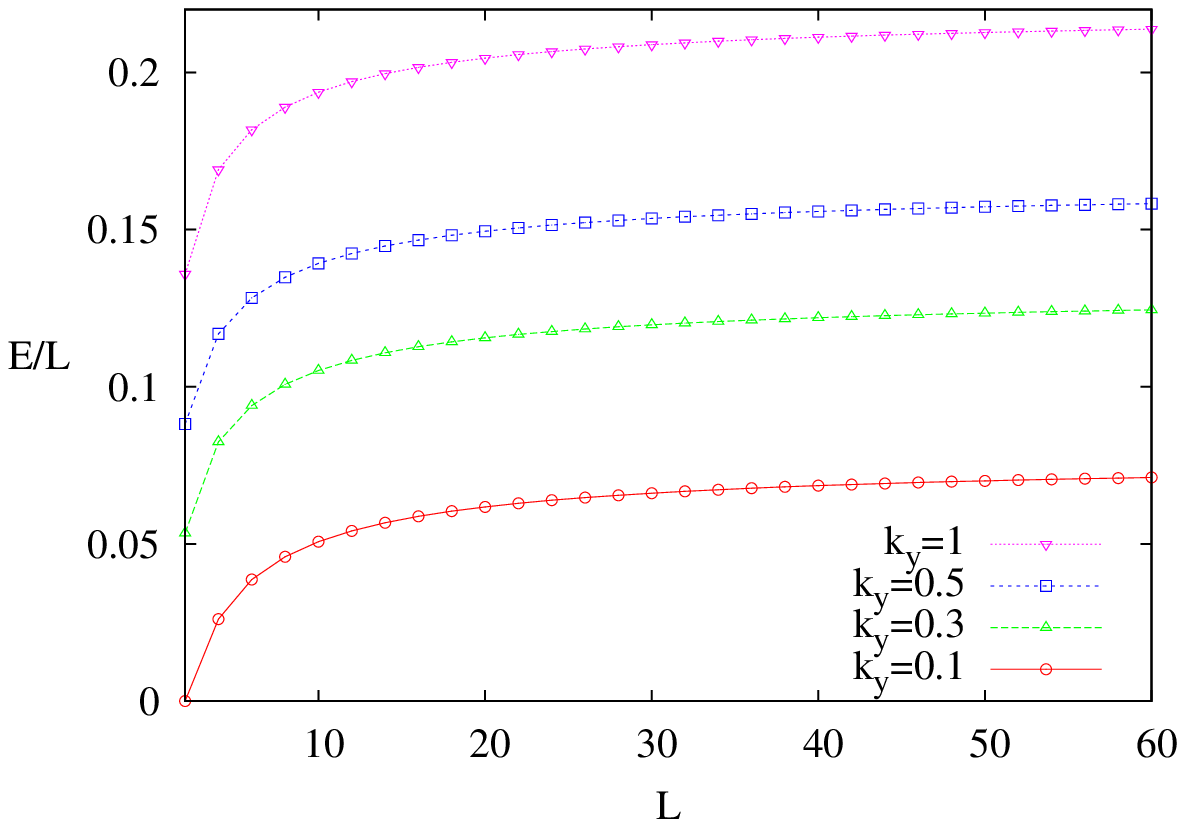}
\caption{Logarithmic negativity per surface area for the harmonic lattice, between two halves
of a $\ell \times \ell$ square with vertical (left) and horizontal (right) partitioning (note the
different vertical scales). The couplings $k_y$ are varied while the other parameters are fixed as
$k_x=1$ and $\Omega_0=10^{-3}$.}
\label{fig:lnosc}
\end{figure*}
%

The Hamiltonian of a 2D lattice of coupled harmonic oscillators is given by
\eq{
H= \frac{1}{2}\sum_n \left( p^2_n + \Omega^2_0 x^2_n  \right)
+ \frac {1}{2} \sum_{\langle m,n \rangle} k_{m,n} (x_m- x_n)^2,
\label{hho}
}
where $\Omega_0$ gives the strength of the harmonic confining potential at each site,
whereas the oscillators are coupled through spring constants $k_{m,n}$ and the sum runs
over all pairs $\langle m,n\rangle$. Here we will restrict ourselves to the case of nearest
neighbor couplings
\eq{
k_{m,n} =
\begin{cases}
k_x & \mbox{if $|i_m-i_n|=1$ and $j_m=j_n$}, \\
k_y & \mbox{if $|j_m-j_n|=1$ and $i_m=i_n$} ,\\
0 & \mbox{otherwise},
\end{cases}
\label{kmn}
}
where the lattice sites are indexed by the pairs of  integers ${\bf R}_m=(i_m,j_m)$.
Introducing the notation ${\bf q}=(q_x,q_y)$ for the wave vectors,
the nonzero matrix elements $\Gamma_{2m-1,2n-1}=2X_{m,n}$ and $\Gamma_{2m,2n}=2P_{m,n}$
of the covariance matrix read
\begin{align}
&2X_{m,n} = \int_{-\pi}^{\pi} \frac{\dd q_x}{2\pi} \int_{-\pi}^{\pi} \frac{\dd q_y}{2\pi}
\ee^{i {\bf q}({\bf R}_m-{\bf R}_n)} \frac{1}{\Omega_{\bf q}} \, ,
\label{xmn} \\
&2P_{m,n} = \int_{-\pi}^{\pi} \frac{\dd q_x}{2\pi} \int_{-\pi}^{\pi} \frac{\dd q_y}{2\pi}
\ee^{i {\bf q}({\bf R}_m-{\bf R}_n)} \Omega_{\bf q} \, ,
\label{pmn}
\end{align}
where the dispersion relation is given by
\eq{
\Omega_{\bf q} = \sqrt{\Omega^2_0+2k_x(1-\cos q_x)+2k_y(1-\cos q_y)}\,.
\label{Omq}}

The RDM of subsystem $A$ has a reduced covariance matrix with
nonzero elements given by $2X_A$ and $2P_A$ and the partial transposition
acts only on $P_A \to P^{T_2}_A$ by changing the signs of
the momenta in $A_2$. Due to the vanishing of cross-correlations between
positions and momenta, the symplectic spectrum of the partial transposed covariance matrix
is simply obtained by finding the eigenvalues $\{ \nu_j \}$ of matrix $\sqrt{2X_A 2P^{T_2}_A}$.
In turn, the logarithmic negativity can be evaluated through Eq.~\eqref{lnbg}.

To be able to compare the results to the 1D case, especially to those obtained via CFT,
we will be interested in a critical lattice system, i.e., in the limit $\Omega_0 \to 0$.
Note, however, that one can not explicitly set $\Omega_0=0$ due to a divergence in
the matrix elements in Eq.~\eqref{xmn} caused by the zero-mode of the lattice.
We have thus used $\Omega_0=10^{-3}$ in the calculations, and we observed that
further decreasing $\Omega_0$ has no visible effect on the results.

For simplicity, we consider only the vertical (Fig.~\ref{fig:2d}a) and horizontal
(Fig.~\ref{fig:2d}b) partitions of a $\ell \times \ell$ square into two halves.
The data for $\lneg$ are shown in Fig.~\ref{fig:lnosc} for the two different geometries,
for different values of the vertical coupling $k_y$ and setting $k_x=1$.
For the vertical partitioning (Fig.~\ref{fig:lnosc} left) and in the limit of uncoupled chains
($k_y=0$), one trivially recovers the $c=1$ CFT result $\lneg /\ell \sim 1/4 \ln \ell + const.$
for the logarithmic negativity per area. However, already a small nonzero $k_y$ leads to
a saturation of the curves, and thus to a strict area law of entanglement. This is indeed
expected from the analogous result on the bipartite entanglement entropy,\cite{BCS06}
which originates from the fact that there is only a single gapless mode within the Brillouin
zone, i.e., one has $\Omega_{\bf q}=0$ only for ${\bf q}=(0,0)$ for any $k_y\ne 0$.
Approaching the isotropic lattice $k_y \to 1$, the entanglement also becomes more evenly
spread out in both directions, leading to a decrease (increase) of the logarithmic negativity
across the vertical (horizontal) cut, as shown on the left (right) of Fig.~\ref{fig:lnosc}.

\subsection{Hopping model}

%
\begin{figure*}[t]
\psfrag{-E3/L}[][][.9]{-$\lneg_3/\ell$}
\psfrag{-E4/L}[][][.9]{-$\lneg_4/\ell$}
\psfrag{L}[][][.9]{$\ell$}
\center
\includegraphics[width=.7\columnwidth]{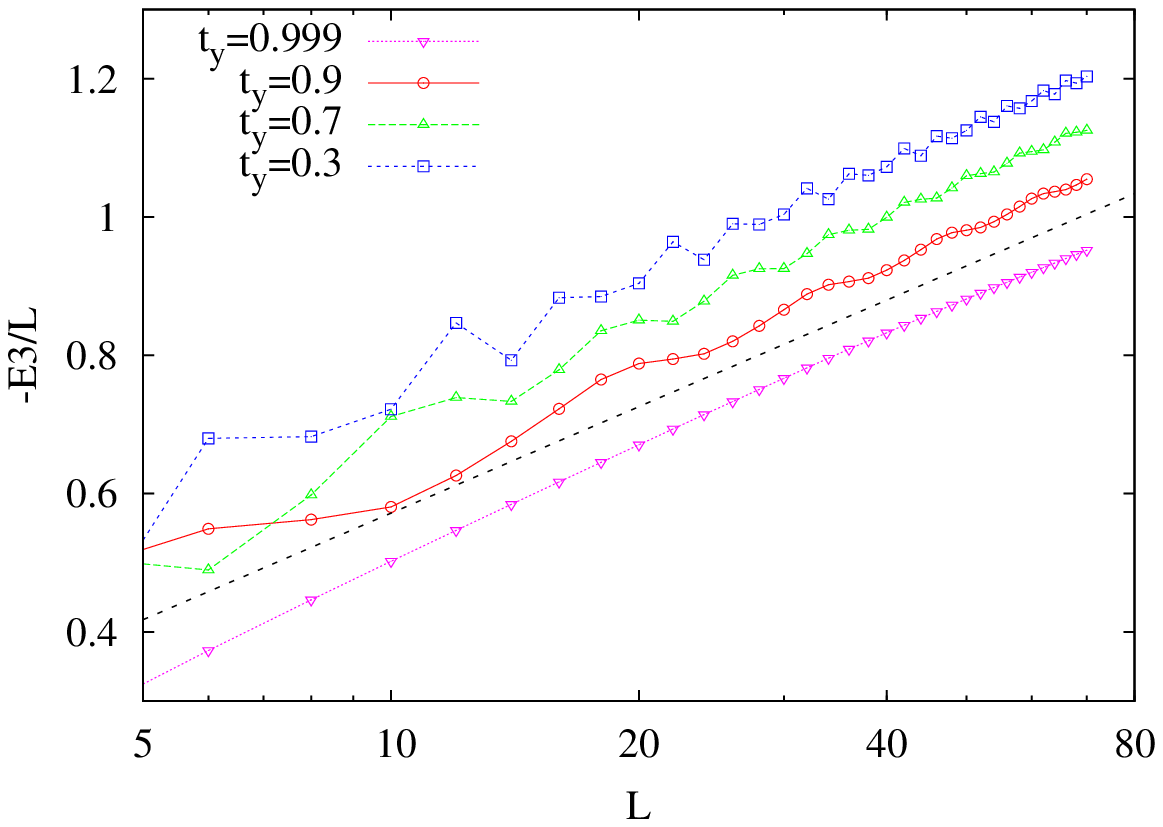}
\hspace{1cm}
\includegraphics[width=.7\columnwidth]{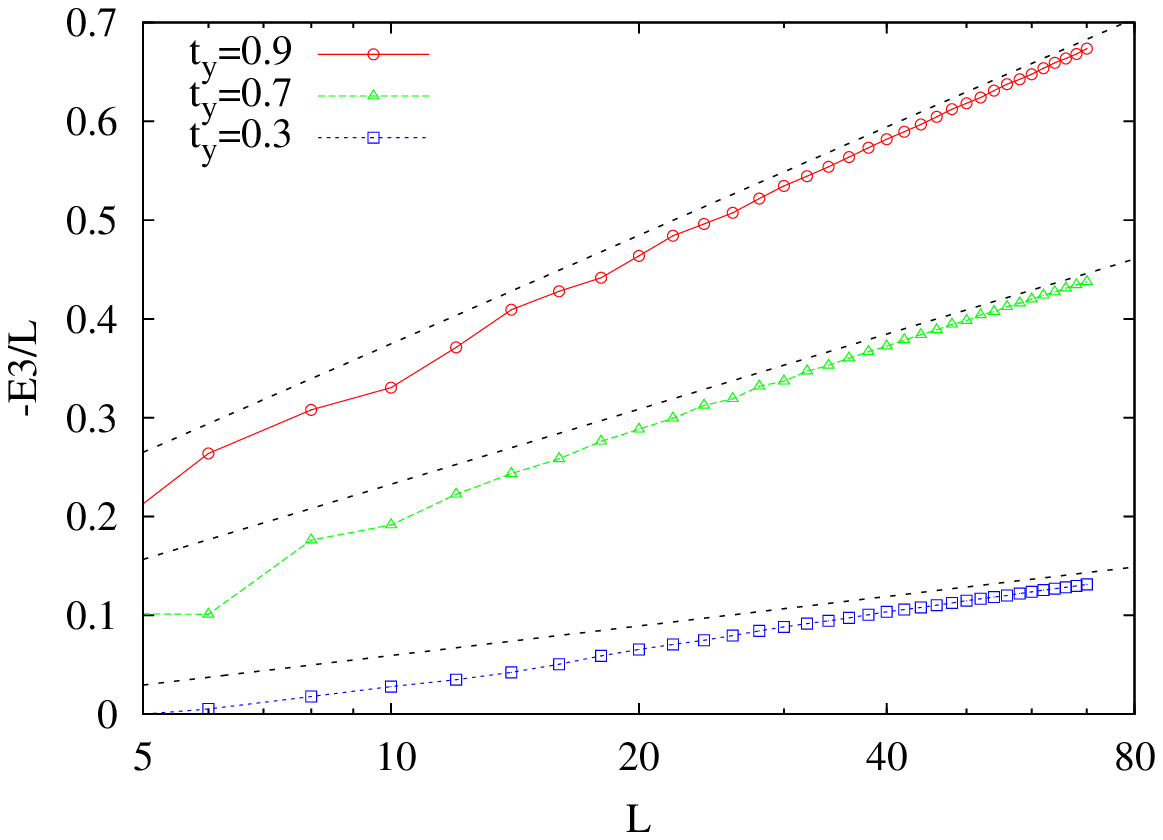}
\includegraphics[width=.7\columnwidth]{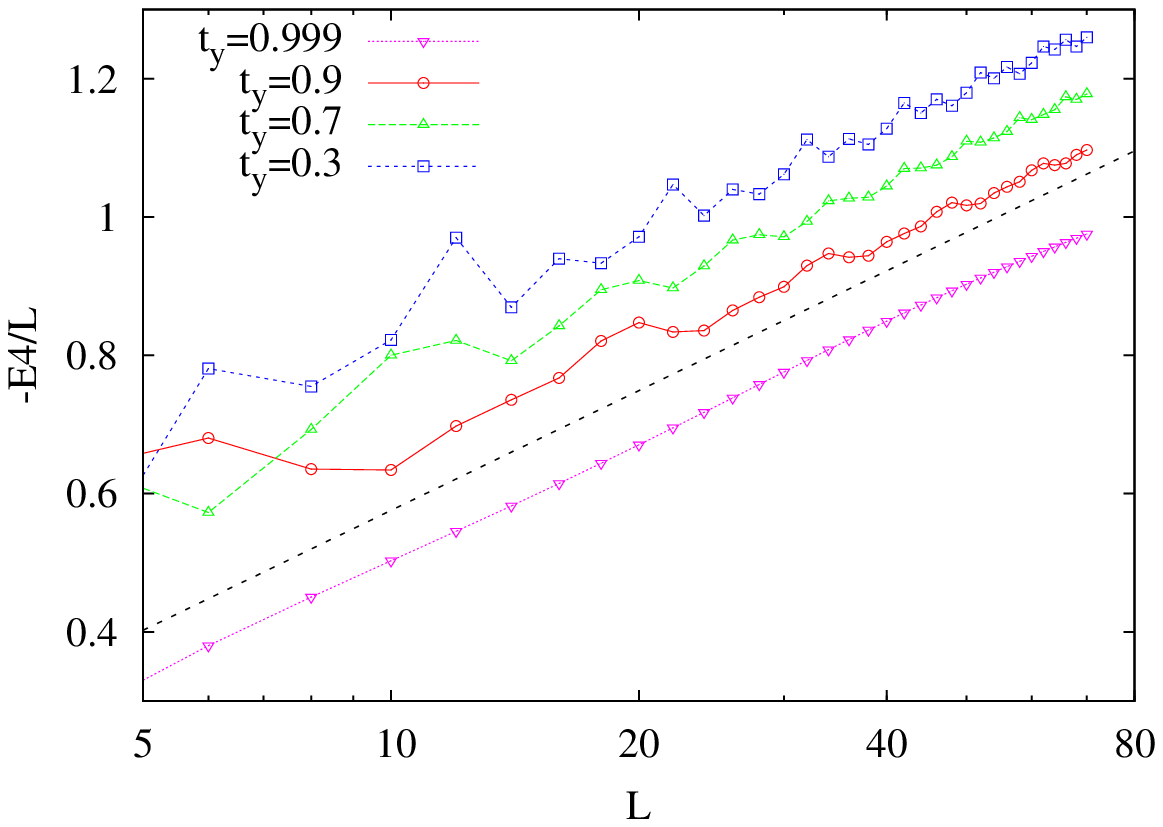}
\hspace{1cm}
\includegraphics[width=.7\columnwidth]{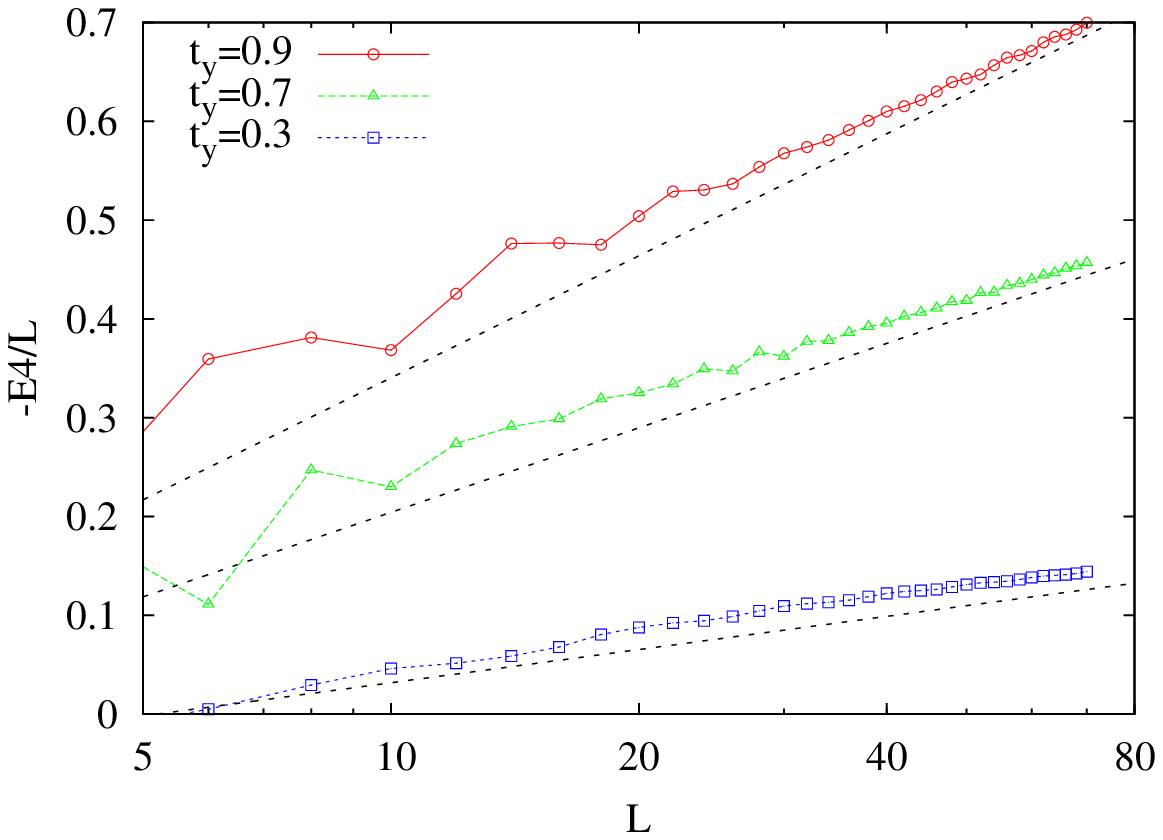}
\caption{Logarithmic ratios per linear size $-\lneg_n/\ell$ for the 2D hopping model with vertical (left) and horizontal (right)
partitioning and various couplings $t_y$. The data is plotted on a logarithmic horizontal scale.
The dashed lines have slopes given by $2/9 \, \sigma_a$ and $2/9 \, \sigma_b$ for $n_o=3$ (up)
and by $1/4 \, \sigma_a$ and $1/4 \, \sigma_b$ for $n_e=4$ (down), respectively, see Eq.~\eqref{sab}.}
\label{fig:enab}
\end{figure*}
%

The planar fermion hopping model is described by the Hamiltonian
\eq{
H=-\frac{1}{2}\sum_{\langle m,n \rangle} t_{m,n} c_m^\dag c_n ,
\label{hff}}
where, analogously to Eq.~\eqref{kmn}, we again consider nearest neighbor hopping only, with amplitudes $t_x$ and $t_y$ in the horizontal and vertical directions, respectively. Since the Hamiltonian given by Eq.~\eqref{hff} is particle-number conserving, the problem simplifies considerably. Indeed, the basic quantities are the correlation functions $C_{m,n}=\langle c^\dag_m c_n \rangle$ which, in the ground state, are given by
\eq{
C_{m,n} = \iint_{{\bf q} \in F} \frac{\dd q_x}{2\pi}\frac{\dd q_y}{2\pi}
\ee^{i {\bf q}({\bf R}_m-{\bf R}_n)}\,.
}
Here the integral goes over the Fermi sea, defined by ${\bf q} \in F$ if $\omega_{\bf q}<0$, with the single-particle dispersion
\eq{
\omega_{\bf q} = -t_x \cos q_x - t_y \cos q_y
\label{omq}
}
obtained through diagonalizing Eq.~\eqref{hff} by a Fourier transform.

Comparing to the formalism introduced in Sec.~\ref{fgs}, one observes
that the following relations hold for the elements of the covariance matrix
\eq{
\begin{split}
&\Gamma_{2m-1,2n}=iG_{m,n}=i(2C_{m,n}-\delta_{m,n}), \\
&\Gamma_{2m,2n-1}=-iG_{m,n}, \quad \Gamma_{2m-1,2n-1}=\Gamma_{2m,2n}=0 \,.
\end{split}
}
Thus, in the presence of particle-number conservation, the spectrum of the reduced
covariance matrix $\Gamma_A$ is directly related to that of the reduced matrix
$G_A=2C_A-\identity$, which in turn determines the eigenvalues of the RDM $\rho_A$.\cite{Peschel03,PE09} In case of the partial transpose $\ptr$, which can be written
as a linear combination of two noncommuting Gaussian operators $O_{A}^{\pm}$ as in Eq.~\eqref{ptrho},
it suffices to consider, instead of $\Gamma_A^\pm$, the spectra of matrices $G_A^{\pm}$
to recover the eigenvalues of $O_{A}^{\pm}$.
Moreover, each moment $\Tr (\ptr)^n$ can be obtained through determinant
formulas involving only $G_A^{\pm}$, as shown in Appendix~\ref{sec:appa}. Note that,
since their linear size is half of the corresponding covariance matrices $\Gamma_A^{\pm}$,
their use is essential to reach large system sizes in our 2D calculations.

To further simplify the setting, we will consider only the geometries, depicted on
Fig.~\ref{fig:2d}, with $A$ being a square of size $\ell \times \ell$ subdivided into
rectangular subsystems $A_1$ and $A_2$ that share a common boundary.
Before presenting our numerical results, and motivated by the results for the entanglement
entropy for 2D free fermions in Eq.~\eqref{eq:dD}, we put forward a conjecture for the behavior
of the logarithmic ratios
\eq{
\lneg_n =
\begin{cases}
-\frac{\sigma}{12}(n_o-1/n_o) \ell \ln \ell, & \mbox{if $n=n_o$ odd}, \\
-\frac{\sigma}{12}(n_e-4/n_e) \ell \ln \ell, & \mbox{if $n=n_e$ even},
\end{cases}
\label{en2d}}
where the geometric factor $\sigma$ is given by
\eq{
\sigma = \frac{1}{4\pi} \int_{\partial F} \dd S_{\bf q}
\int_{\partial A_{12}} \dd S_{\bf r} |\mathbf{n}_{\bf q} \mathbf{n}_{\bf r}|
\label{sigma} \, .
}
Here the momentum and real space integrals have to be carried out along the Fermi
surface $\partial F$ and the common boundary $\partial A_{12}$ of subsystems $A_1$
and $A_2$, respectively, and the linear size $\ell$ has been scaled out such that $A$
becomes a unit square. Note that the prefactor \eqref{sigma} is chosen such that,
in the limit $t_y=0$ of decoupled 1D chains, one recovers $\sigma = c =1$ the central
charge of the free fermion CFT, and Eq.~\eqref{en2d} reproduces the result \eqref{en1dadj}
for the adjacent intervals. Similarly, the doubling of the prefactor $\sigma=2$ and hence
Eq.~\eqref{en1demb} is recovered for the 1D embedded geometry.
For the particular choices of the 2D partitions shown on Fig.~\ref{fig:2d} a) and b),
a simple calculation (see Appendix~\ref{sec:appb}) yields
\eq{
\sigma_a = 1, \quad \sigma_b = \frac{2}{\pi} \arcsin(t_y),
\label{sab}
}
whereas for cases c) and d) one trivially finds
\eq{
\sigma_c = (\sigma_a+\sigma_b)/2, \quad \sigma_d=2\sigma_c .
\label{scd}
}
%

%
\begin{figure*}[t]
\psfrag{-E3/L}[][][.9]{-$\lneg_3/\ell$}
\psfrag{-E4/L}[][][.9]{-$\lneg_4/\ell$}
\psfrag{L}[][][.9]{$\ell$}
\center
\includegraphics[width=.7\columnwidth]{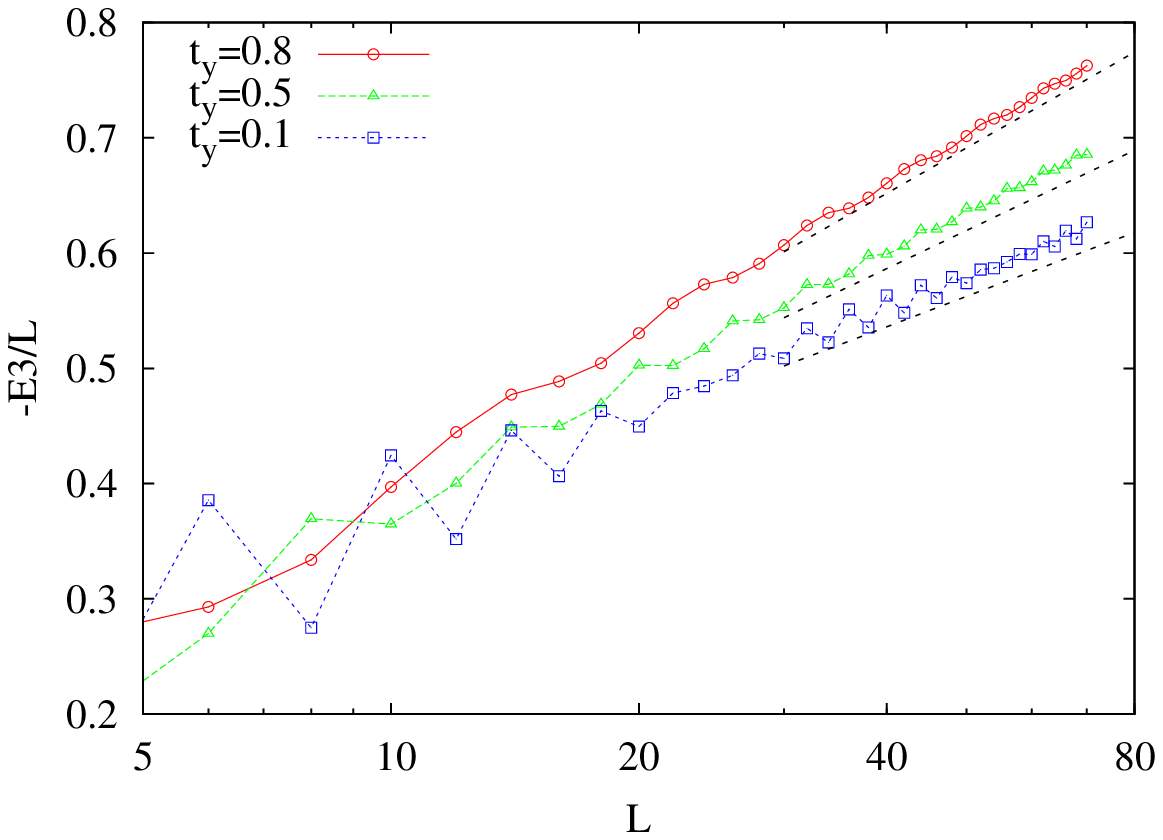}
\hspace{1cm}
\includegraphics[width=.7\columnwidth]{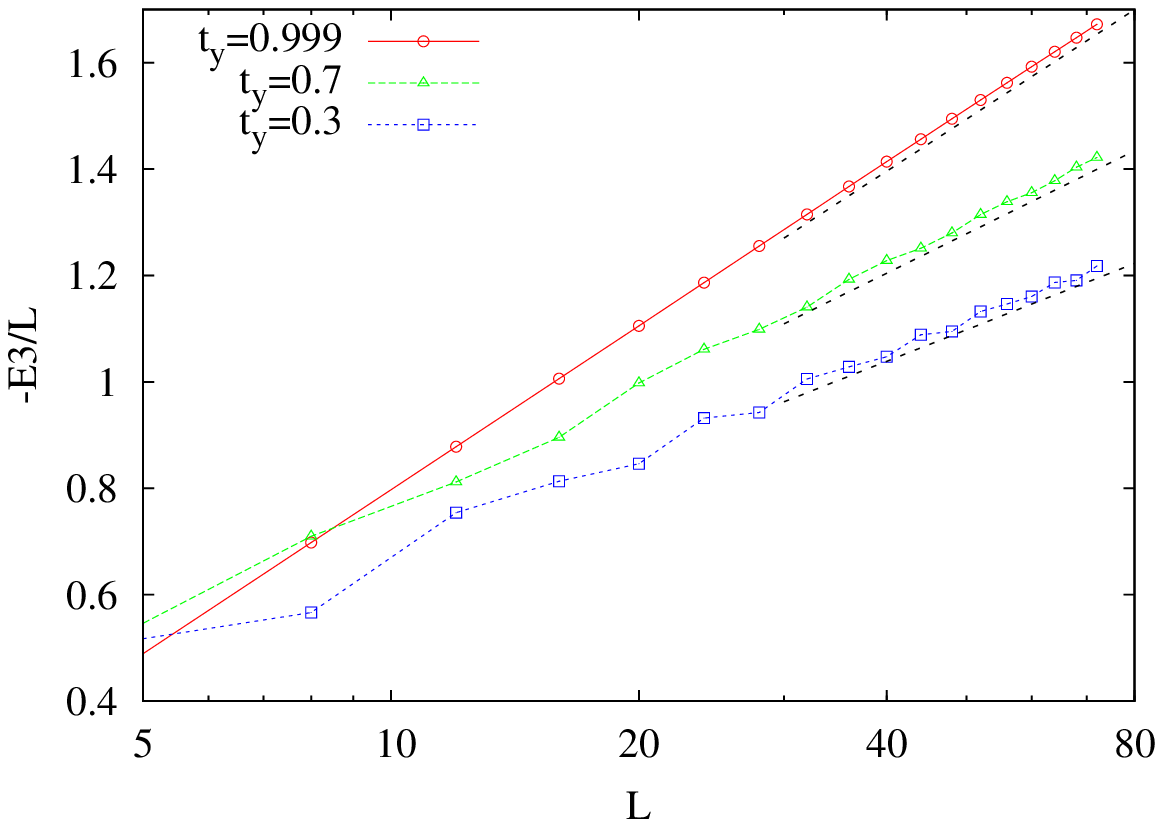}
\includegraphics[width=.7\columnwidth]{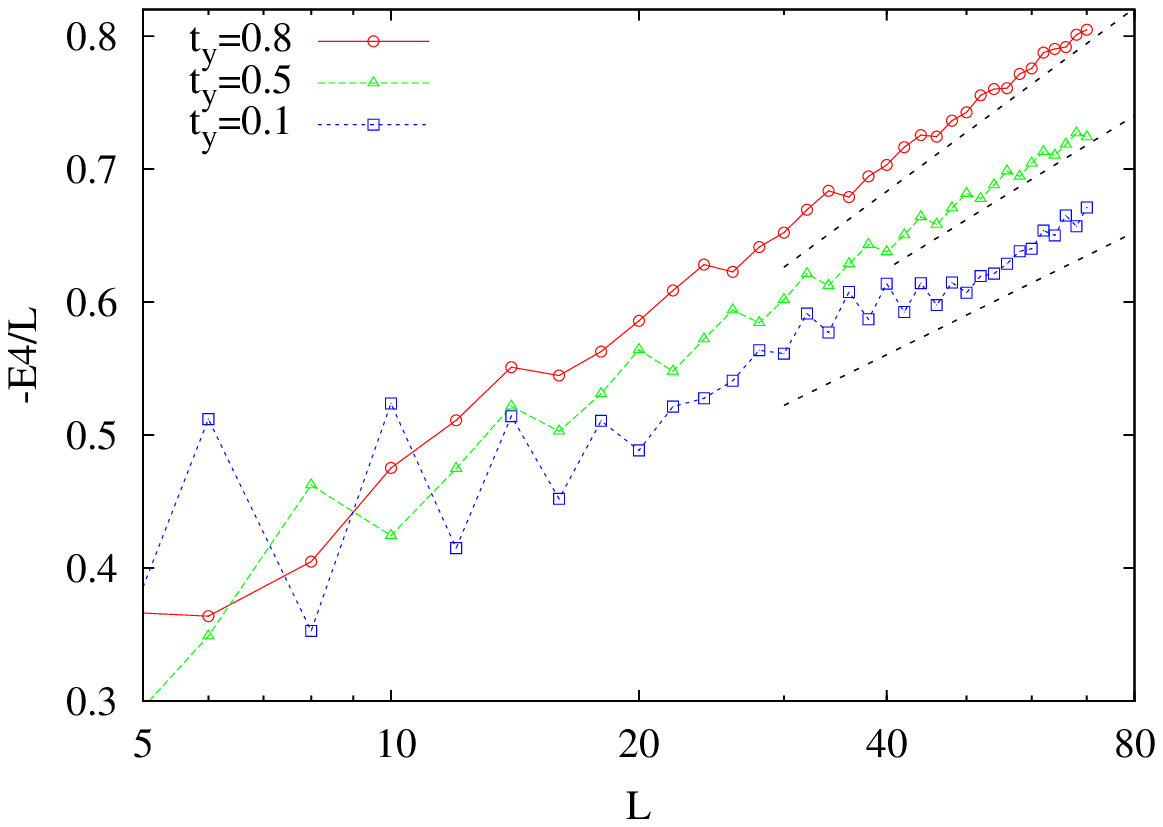}
\hspace{1cm}
\includegraphics[width=.7\columnwidth]{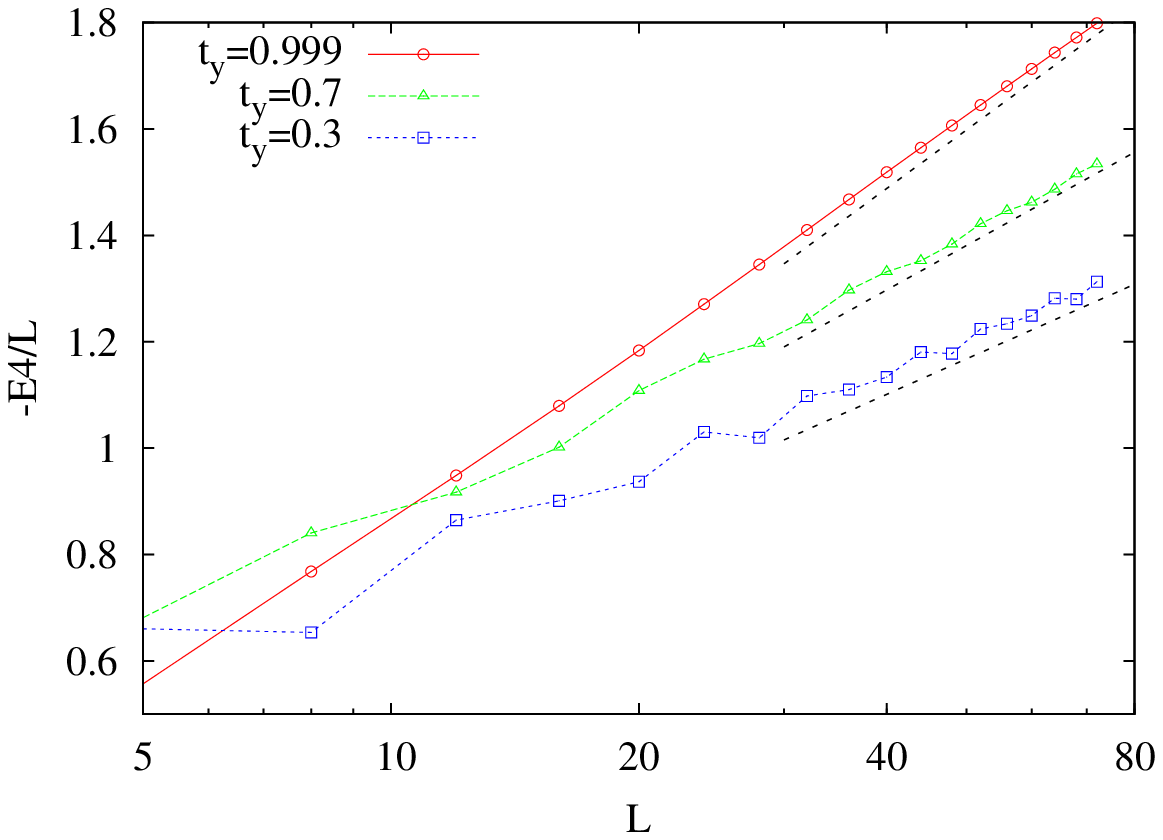}
\caption{Logarithmic ratios per linear size $-\lneg_n/\ell$ for the 2D hopping model with corner (left) and embedded (right)
partitioning and various couplings $t_y$. The data is plotted on a logarithmic horizontal scale.
The dashed lines have slopes given by $2/9 \, \sigma_c$ and $2/9 \, \sigma_d$ for $n_o=3$ (up)
and by $1/4 \, \sigma_c$ and $1/4 \, \sigma_d$ for $n_e=4$ (down), respectively, see Eq.~\eqref{scd}.}
\label{fig:encd}
\end{figure*}
%

The numerical results for $\lneg_n$ with $n=3$ and $n=4$, obtained by evaluating the
determinant formulas in Appendix~\ref{sec:appa}, are shown in Figs. \ref{fig:enab}
and \ref{fig:encd}. In these figures we plotted $-\lneg_n/\ell$ on a logarithmic scale,
to best visualize the expected behavior \eqref{en2d}. The maximal linear size $\ell=70$
reached in our calculations is unfortunately still rather small to extract the slopes of the curves
through fitting, as they become unstable due to the presence of subleading corrections.
Instead, we simply plot the conjectured slope of the curves with dashed lines and compare it to the data sets. 

In case of the vertical partitioning, one expects a slope which is independent
of the vertical hopping amplitude $t_y$. This is indeed nicely recovered from our data, shown in the
left of Fig.~\ref{fig:enab}, and even the numerical value of the slope fits very well to our conjecture.
Decreasing $t_y$, the area-law contribution increases and the curves are shifted upwards, which
is the same trend observed for the negativity of the harmonic lattice, see left of Fig.~\ref{fig:lnosc}.
Additionally, the curves pick up oscillatory contributions with an increasing frequency.

On the other hand, for a horizontal partitioning the geometric prefactor $\sigma_b$ in \eqref{sab}
is a decreasing function of $t_y$ and thus the slope of the curves should tend to zero for $t_y \to 0$, 
i.e., when the partition becomes disconnected. This is clearly observed from
our numerical data in the right of Fig.~\ref{fig:enab}. In this case, however, it is somewhat more
difficult to conclude about the correctness of our conjecture, as the data shows significant subleading
corrections up to the reachable system sizes. Nevertheless, for $n_o=3$ one still finds a good agreement
which, however, deteriorates for $n_e=4$. In fact, for 1D systems, the presence of unusual corrections
whose magnitude increases with the index is well known for the R\'enyi entropies from CFT calculations
\cite{CC10} and was also observed for the moments of the partial transpose.\cite{CTC15}

The data for the corner and the embedded partitionings are shown on the left and right of
Fig.~\ref{fig:encd}, respectively. Due to its geometric nature, the corner prefactor $\sigma_c$
is just the average of the vertical and horizontal ones, whereas the prefactor for the
embedded geometry $\sigma_d$ is the double of the corner prefactor. This is indeed
in very good agreement with the numerical data, especially for the embedded geometry
which seems to be the least effected by subleading corrections.

\section{Conclusions}\label{sec:Discussion}

We have studied the scaling of entanglement negativity in ground states of 2D free lattice systems between rectangular regions having a common boundary. While for harmonic oscillators a strict area law is obeyed, we found logarithmic corrections for the moments of the partial transpose in the fermionic case, which is completely analogous to the result for bipartite entanglement entropies. Based on this similarity and on CFT results for 1D systems, we conjectured a geometric form \eqref{sigma} for the prefactor governing the leading behavior \eqref{en2d} of the logarithmic ratios for the planar fermionic hopping model, and a comparison with numerical calculations shows a good agreement.

It would be interesting to find a strict proof for the form of the area-law violating term which, in the case of bipartite entanglement entropies, is related to the Widom conjecture\cite{GK06} and has only been proved recently.\cite{LSS14} In contrast, for the moments of the partial transpose we do not even have an analogue of the method of Ref.~[\onlinecite{JK04}] for 1D free fermions, where R\'enyi entropies are calculated using the asymptotics of Toeplitz determinants. Although $\Tr (\ptr)^n$ can also be cast as a sum of determinants, these are not of the Toeplitz type and we have not yet been able to find their asymptotics analytically. This would clearly be a necessary first step in order to understand the 2D results, and thus requires further studies.

There are a number of possible extensions of the set-up presented in our work. Firstly, it would be interesting to see whether an interacting 2D Fermi liquid would show a similar negativity scaling as free fermions, which one would expect from the simple physical picture discussed in Section~\ref{sec:2d}. The recent advances in numerical methods for evaluating entanglement negativity for interacting systems\cite{WMB09,Bayat10,CTT13,Chung14, SDHS15} cast some hope that this could be answered in the near future. Secondly, it would also be natural to investigate how the logarithmic correction to the negativity area law is rounded off when the Fermi surface degenerates to a number of points, and also to study corner effects. Thirdly, the question how the negativity decays with distance between non-adjacent subsystems should also be addressed. Finally, the possibility of detecting topological order via negativity\cite{LV13,Castel13} for free-fermion systems, using the Gaussian toolbox presented here, is left for future study.

\begin{acknowledgments}

We thank A.~Coser, C.~De~Nobili, E.~Tonni and J.~Eisert for helpful discussions. V.E. acknowledges funding from the Austrian Science Fund (FWF) through Lise Meitner Project No. M1854-N36. The work of Z.Z. was supported by the British Engineering and Physical Sciences Research Council (EPSRC).

\end{acknowledgments}

\appendix

\section{Trace formulas\label{sec:appa}}

Here we give the necessary formulae to evaluate the moments $\Tr (\ptr)^n$
of the partial transpose in the ground state of a particle-number conserving free-fermion
Hamiltonian. It will be additionally assumed that the correlation matrix $C_{m,n}=\langle c^\dag_m c_n \rangle$ is real, which holds for the models studied in the paper.

As described in the main text, the partial transpose can be given as a linear
combination of two Gaussian operators, see Eq.~\eqref{ptrho}. To simplify notation,
we shall omit the subscripts $A$ here and use $O_\pm$. In terms of the fermionic operators
$c^\dag_k$ and $c_k$ they are given by the quadratic form
\eq{
O_\sigma = \frac{1}{Z_\sigma}\exp \left( \sum_{k,l} (W_\sigma)_{k,l} c^\dag_k c_l \right),
\label{opm}}
where $\sigma=\pm$ and $Z_\sigma$ ensures normalization, $\Tr O_\sigma$=1.
The matrices in the exponent satisfy
\eq{
\tanh \frac{W_\sigma}{2} = G_\sigma, \qquad
\exp (W_\sigma) = \frac{1+G_\sigma}{1-G_\sigma} \, ,
\label{wg}}
where $G_\sigma$ is defined through the correlation matrix $C$ as
\eq{
G_\sigma = \twomat{(2C-\identity)_{11}}{\sigma i(2C-\identity)_{12}}
{\sigma i (2C-\identity)_{21}}{-(2C-\identity)_{22}} \, ,
}
and the subscripts refer to the reduction of matrices (rows and columns, respectively)
to the corresponding subsystems $A_1$ and $A_2$.

To obtain the $n$-th moment, one has first to factor out Eq.~\eqref{ptrho}, which yields
\eq{
\Tr (\ptr)^n = \sum_{\{\sigma_i\}} \frac{\exp(-i\frac{\pi}{4}\sum_{i=1}^{n}\sigma_i)}{2^{n/2}}
\Tr(\prod_{i=1}^{n} O_{\sigma_{i}}).
\label{ptrfac}}
Note that the sum goes over all the possible assignments of $\{\sigma_i\}$ and
$\sum_{i} \sigma_i$ is just the difference between the numbers of $+$ and $-$
terms in the corresponding factor. Using the fact that each $O_{\sigma_i}$ is a Gaussian
operator given by Eq.~\eqref{opm}, one can apply determinant formulas for the traces
of their products. Indeed, using the product relation for general Gaussian operators
\eq{
\prod_{i=1}^n \exp\left(\textstyle{\sum_{k,l}} (W_{\sigma_i} )_{k,l} c^\dag_k c_l  \right)
= \exp\left(\textstyle{\sum_{k,l}} V_{k,l} c^\dag_k c_l \right) ,
}
where $\exp(V)= \prod_{i=1}^n \exp(W_{\sigma_i})$, and the trace formula\cite{Klich02}
\eq{
\Tr\left( \exp ( \textstyle{\sum_{k,l}} V_{k,l} c^\dag_k c_l )\right)=\det \left[ \identity + \exp(V) \right] ,
} 
one obtains
\eq{
\Tr(\prod_{i=1}^{n} O_{\sigma_{i}}) =
\frac{\det \left[ \identity + \prod_{i=1}^{n}\exp(W_{\sigma_i}) \right]}
{\prod_{i=1}^{n} \det \left[ \identity+\exp(W_{\sigma_i}) \right]} \, ,
\label{trdet}}
where the denominator is just the normalization factor $\prod_{i=1}^{n}Z_{\sigma_i}$.
Hence, the result \eqref{ptrfac} is given by a weighted sum of $2^n$ determinants. In fact, it was recently
pointed out that, for free fermions in 1D, each of these terms can be associated to partition
functions on a higher genus Riemann surface with different \emph{spin structures}.\cite{CTC15b}

Using the property $O_- = (O_+)^\dag$ and thus the invariance of the formula under
the exchange $\sigma_i \to -\sigma_i$, the number of determinants to be evaluated
can be reduced. Furthermore, using the relation \eqref{wg} between
matrices $W_\sigma$ and $G_\sigma$, each determinant in \eqref{trdet} can be rewritten
in terms of the latter ones. In particular, for $n=3$ and $n=4$ we obtain, after simple
but lengthy algebra, the following expressions
\eq{
\Tr(O_+^3) =
\det \left(\frac{\identity+3G_+^2}{4}\right) ,
}
\eq{
\Tr(O_+^2O_-) =
\det \left(\frac{\identity+G_+^2+2G_+G_-}{4}\right)  ,
}
\eq{
\Tr(O_+^4) =
\det \left(\frac{\identity+6G_+^2+G_+^4}{8}\right)  ,
}
\eq{
\Tr(O_+^2O_-^2) =
\det \left(\frac{(\identity+G_+^2)(\identity+G_-^2)+4G_+G_-}{8}\right)  ,
}
\eq{
\Tr(O_+O_-O_+O_-) =
\det \left( \frac{(\identity+G_{+}G_-)^2 \left( \identity{+}G_{\times}^2 \right)}{8}\right)  ,
}
where
\eq{
G_\times = \identity - (\identity-G_-)(\identity+G_+G_-)^{-1}(\identity-G_+) \, .
}

Finally, one should note that very similar formulas can be applied
in the general free-fermion case without particle-number conservation. \cite{EZ15}
Then the matrices $G_\sigma$ should be exchanged with the modified covariance matrices
$\Gamma_\sigma$, given in Eq.~\eqref{covf}, and one has to take the square root of the determinants.
Note, however, that the resulting expressions then involve a sign ambiguity \cite{FC10} which has its root
in the underlying Pfaffian structure. \cite{Klich14}

\section{Calculation of the geometric prefactor\label{sec:appb}}

In this appendix we show how to calculate the geometric prefactor $\sigma$
which determines the slope of the curves in Figs.~\ref{fig:enab} and \ref{fig:encd},
using a method very similar to the one presented in Ref.~[\onlinecite{RSLHS13}].

%
\begin{figure}[htb]
\center
\includegraphics[width=.9\columnwidth]{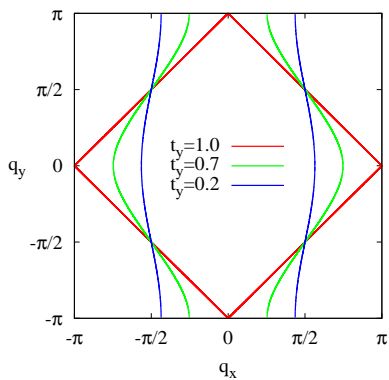}
\caption{The Fermi surface of the infinite 2D hopping model for various anisotropies $t_y$ and $t_x=1$.}
\label{fig:fs}
\end{figure}
%

According to Eq.~\eqref{sigma}, the prefactor is given by a double integral over the
Fermi surface $\partial F$ and over the surface $\partial A_{12}$ separating subsystems
$A_1$ and $A_2$. The Fermi surface is defined through the single-particle dispersion
\eqref{omq} as ${\bf q} \in \partial F$ if $\omega_{\bf q} = 0$
and is depicted on Fig.~\ref{fig:fs} for various anisotropies. Note that the Fermi surface
is invariant under reflections $q_x \to -q_x$ and $q_y \to -q_y$, it is thus enough to treat
the first quadrant $\partial F_1$ in the integral \eqref{sigma} and multiply the result by four.
Furthermore, for the geometries shown in Fig.~\ref{fig:2d} a) and b), the normal vector
on the entire real-space surface $\partial A_{12}$ is constant and given by ${\bf n_r}=(1,0)$
and ${\bf n_r}=(0,1)$, respectively. Thus, in these cases the real-space integration can
be dropped, whereas for the geometries in Fig.~\ref{fig:2d} c) and d) the results can be 
obtained trivially by combining those of a) and b).

The normal vector along the Fermi surface is given by
\eq{
{\bf n_q} = \frac{\nabla_{\bf q}\omega_{\bf q}}{|\nabla_{\bf q}\omega_{\bf q}|} \, ,
\label{nq}}
and the path of integration can be parametrized as $q_x(\theta)$, $q_y(\theta)$,
with the line element given by
\eq{
\dd S_{\bf q} = \sqrt{\left(\frac{\dd q_x}{\dd \theta}\right)^2 +
\left(\frac{\dd q_y}{\dd \theta}\right)^2} \dd \theta \, .
\label{daq}}
Furthermore, we can use the fact that the dispersion is constant (i.e. zero) along the Fermi surface
\eq{
\frac{\partial \omega_{\bf q}}{\partial \theta} =
\frac{\partial \omega_{\bf q}}{\partial q_x} \frac{\dd q_x}{\dd \theta} +
\frac{\partial \omega_{\bf q}}{\partial q_y} \frac{\dd q_y}{\dd \theta} = 0 \, ,
}
which can be used to relate the line element in \eqref{daq} to the denominator
of the normal vector in \eqref{nq}. After proper cancellations, one finds the simple results
\eq{
\sigma_a = \int_{\partial F_1} \frac{\dd \theta}{\pi} \left| \frac{\dd q_y}{\dd \theta} \right|,
\quad
\sigma_b = \int_{\partial F_1} \frac{\dd \theta}{\pi} \left| \frac{\dd q_x}{\dd \theta} \right| \, .
\label{sab2}}

Thus, the prefactors are simply related to the overall change in the components of the wavenumber
as one sweeps along the Fermi surface. Looking at Fig.~\ref{fig:fs}, the change in $q_y$ within the
first quadrant is always given by $\pi$, independent of $t_y$. On the other hand, the extent of the
Fermi surface in the $x$-direction shrinks for larger anisotropies, with the locations of  the endpoints
given by  $q_x=\pi/2 \pm \arcsin(t_y)$. Substituting into Eq.~\eqref{sab2}, one obtains the results in \eqref{sab}.

\bibliographystyle{apsrev.bst}

\bibliography{negat2d_refs}

\begin{thebibliography}{68}
\expandafter\ifx\csname natexlab\endcsname\relax\def\natexlab#1{#1}\fi
\expandafter\ifx\csname bibnamefont\endcsname\relax
  \def\bibnamefont#1{#1}\fi
\expandafter\ifx\csname bibfnamefont\endcsname\relax
  \def\bibfnamefont#1{#1}\fi
\expandafter\ifx\csname citenamefont\endcsname\relax
  \def\citenamefont#1{#1}\fi
\expandafter\ifx\csname url\endcsname\relax
  \def\url#1{\texttt{#1}}\fi
\expandafter\ifx\csname urlprefix\endcsname\relax\def\urlprefix{URL }\fi
\providecommand{\bibinfo}[2]{#2}
\providecommand{\eprint}[2][]{\url{#2}}

\bibitem[{\citenamefont{Eisert et~al.}(2010)\citenamefont{Eisert, Cramer, and
  Plenio}}]{ECP09}
\bibinfo{author}{\bibfnamefont{J.}~\bibnamefont{Eisert}},
  \bibinfo{author}{\bibfnamefont{M.}~\bibnamefont{Cramer}}, \bibnamefont{and}
  \bibinfo{author}{\bibfnamefont{M.~B.} \bibnamefont{Plenio}},
  \bibinfo{journal}{Rev. Mod. Phys.} \textbf{\bibinfo{volume}{82}},
  \bibinfo{pages}{277} (\bibinfo{year}{2010}).

\bibitem[{\citenamefont{Calabrese et~al.}(2009)\citenamefont{Calabrese, Cardy,
  and Doyon}}]{CCD09}
\bibinfo{author}{\bibfnamefont{P.}~\bibnamefont{Calabrese}},
  \bibinfo{author}{\bibfnamefont{J.}~\bibnamefont{Cardy}}, \bibnamefont{and}
  \bibinfo{author}{\bibfnamefont{B.}~\bibnamefont{Doyon}}, \bibinfo{journal}{J.
  Phys. A: Math. Theor.} \textbf{\bibinfo{volume}{42}}, \bibinfo{pages}{500301}
  (\bibinfo{year}{2009}).

\bibitem[{\citenamefont{Vidal et~al.}(2003)\citenamefont{Vidal, Latorre, Rico,
  and Kitaev}}]{Vidal03}
\bibinfo{author}{\bibfnamefont{G.}~\bibnamefont{Vidal}},
  \bibinfo{author}{\bibfnamefont{J.~I.} \bibnamefont{Latorre}},
  \bibinfo{author}{\bibfnamefont{E.}~\bibnamefont{Rico}}, \bibnamefont{and}
  \bibinfo{author}{\bibfnamefont{A.}~\bibnamefont{Kitaev}},
  \bibinfo{journal}{Phys. Rev. Lett.} \textbf{\bibinfo{volume}{90}},
  \bibinfo{pages}{227902} (\bibinfo{year}{2003}).

\bibitem[{\citenamefont{Calabrese and Cardy}(2004)}]{CC04}
\bibinfo{author}{\bibfnamefont{P.}~\bibnamefont{Calabrese}} \bibnamefont{and}
  \bibinfo{author}{\bibfnamefont{J.~L.} \bibnamefont{Cardy}},
  \bibinfo{journal}{J. Stat. Mech.} \bibinfo{eid}{P06002}
  (\bibinfo{year}{2004}).

\bibitem[{\citenamefont{Kitaev and Preskill}(2006)}]{Kitaev06}
\bibinfo{author}{\bibfnamefont{A.}~\bibnamefont{Kitaev}} \bibnamefont{and}
  \bibinfo{author}{\bibfnamefont{J.}~\bibnamefont{Preskill}},
  \bibinfo{journal}{Phys. Rev. Lett.} \textbf{\bibinfo{volume}{96}},
  \bibinfo{pages}{110404} (\bibinfo{year}{2006}).

\bibitem[{\citenamefont{Levin and Wen}(2006)}]{LevinWen06}
\bibinfo{author}{\bibfnamefont{M.}~\bibnamefont{Levin}} \bibnamefont{and}
  \bibinfo{author}{\bibfnamefont{X.-G.} \bibnamefont{Wen}},
  \bibinfo{journal}{Phys. Rev. Lett.} \textbf{\bibinfo{volume}{96}},
  \bibinfo{pages}{110405} (\bibinfo{year}{2006}).

\bibitem[{\citenamefont{Hamma et~al.}(2005)\citenamefont{Hamma, Ionicioiu, and
  Zanardi}}]{Hamma05}
\bibinfo{author}{\bibfnamefont{A.}~\bibnamefont{Hamma}},
  \bibinfo{author}{\bibfnamefont{R.}~\bibnamefont{Ionicioiu}},
  \bibnamefont{and} \bibinfo{author}{\bibfnamefont{P.}~\bibnamefont{Zanardi}},
  \bibinfo{journal}{Phys. Lett. A} \textbf{\bibinfo{volume}{337}},
  \bibinfo{pages}{22} (\bibinfo{year}{2005}).

\bibitem[{\citenamefont{Lee et~al.}(2000)\citenamefont{Lee, Kim, Park, and
  Lee}}]{Kim00}
\bibinfo{author}{\bibfnamefont{J.}~\bibnamefont{Lee}},
  \bibinfo{author}{\bibfnamefont{M.~S.} \bibnamefont{Kim}},
  \bibinfo{author}{\bibfnamefont{Y.~J.} \bibnamefont{Park}}, \bibnamefont{and}
  \bibinfo{author}{\bibfnamefont{S.}~\bibnamefont{Lee}}, \bibinfo{journal}{J.
  Mod. Opt.} \textbf{\bibinfo{volume}{47}}, \bibinfo{pages}{2151}
  (\bibinfo{year}{2000}).

\bibitem[{\citenamefont{Vidal and Werner}(2002)}]{VW02}
\bibinfo{author}{\bibfnamefont{G.}~\bibnamefont{Vidal}} \bibnamefont{and}
  \bibinfo{author}{\bibfnamefont{R.~F.} \bibnamefont{Werner}},
  \bibinfo{journal}{Phys. Rev. A} \textbf{\bibinfo{volume}{65}},
  \bibinfo{pages}{032314} (\bibinfo{year}{2002}).

\bibitem[{\citenamefont{Plenio}(2005)}]{Plenio05}
\bibinfo{author}{\bibfnamefont{M.~B.} \bibnamefont{Plenio}},
  \bibinfo{journal}{Phys. Rev. Lett.} \textbf{\bibinfo{volume}{95}},
  \bibinfo{pages}{090503} (\bibinfo{year}{2005}).

\bibitem[{\citenamefont{Simon}(2000)}]{Simon00}
\bibinfo{author}{\bibfnamefont{R.}~\bibnamefont{Simon}},
  \bibinfo{journal}{Phys. Rev. Lett.} \textbf{\bibinfo{volume}{84}},
  \bibinfo{pages}{2726} (\bibinfo{year}{2000}).

\bibitem[{\citenamefont{Audenaert et~al.}(2002)\citenamefont{Audenaert, Eisert,
  Plenio, and Werner}}]{AEPW02}
\bibinfo{author}{\bibfnamefont{K.}~\bibnamefont{Audenaert}},
  \bibinfo{author}{\bibfnamefont{J.}~\bibnamefont{Eisert}},
  \bibinfo{author}{\bibfnamefont{M.~B.} \bibnamefont{Plenio}},
  \bibnamefont{and} \bibinfo{author}{\bibfnamefont{R.~F.}
  \bibnamefont{Werner}}, \bibinfo{journal}{Phys. Rev. A}
  \textbf{\bibinfo{volume}{66}}, \bibinfo{pages}{042327}
  (\bibinfo{year}{2002}).

\bibitem[{\citenamefont{Eisler and Zimbor\'as}(2015)}]{EZ15}
\bibinfo{author}{\bibfnamefont{V.}~\bibnamefont{Eisler}} \bibnamefont{and}
  \bibinfo{author}{\bibfnamefont{Z.}~\bibnamefont{Zimbor\'as}},
  \bibinfo{journal}{New J. Phys.} \textbf{\bibinfo{volume}{17}},
  \bibinfo{pages}{053048} (\bibinfo{year}{2015}).

\bibitem[{\citenamefont{Calabrese
  et~al.}(2012{\natexlab{a}})\citenamefont{Calabrese, Cardy, and
  Tonni}}]{CCT12}
\bibinfo{author}{\bibfnamefont{P.}~\bibnamefont{Calabrese}},
  \bibinfo{author}{\bibfnamefont{J.}~\bibnamefont{Cardy}}, \bibnamefont{and}
  \bibinfo{author}{\bibfnamefont{E.}~\bibnamefont{Tonni}},
  \bibinfo{journal}{Phys. Rev. Lett.} \textbf{\bibinfo{volume}{109}},
  \bibinfo{pages}{130502} (\bibinfo{year}{2012}{\natexlab{a}}).

\bibitem[{\citenamefont{Calabrese
  et~al.}(2013{\natexlab{a}})\citenamefont{Calabrese, Cardy, and
  Tonni}}]{CCT13}
\bibinfo{author}{\bibfnamefont{P.}~\bibnamefont{Calabrese}},
  \bibinfo{author}{\bibfnamefont{J.}~\bibnamefont{Cardy}}, \bibnamefont{and}
  \bibinfo{author}{\bibfnamefont{E.}~\bibnamefont{Tonni}}, \bibinfo{journal}{J.
  Stat. Mech.} \bibinfo{eid}{P02008} (\bibinfo{year}{2013}{\natexlab{a}}).

\bibitem[{\citenamefont{\relax{De~Nobili}
  et~al.}(2015)\citenamefont{\relax{De~Nobili}, Coser, and Tonni}}]{NCT15}
\bibinfo{author}{\bibfnamefont{C.}~\bibnamefont{\relax{De~Nobili}}},
  \bibinfo{author}{\bibfnamefont{A.}~\bibnamefont{Coser}}, \bibnamefont{and}
  \bibinfo{author}{\bibfnamefont{E.}~\bibnamefont{Tonni}}, \bibinfo{journal}{J.
  Stat. Mech.} \bibinfo{eid}{P06021} (\bibinfo{year}{2015}).

\bibitem[{\citenamefont{Coser et~al.}(2015)\citenamefont{Coser, Tonni, and
  Calabrese}}]{CTC15}
\bibinfo{author}{\bibfnamefont{A.}~\bibnamefont{Coser}},
  \bibinfo{author}{\bibfnamefont{E.}~\bibnamefont{Tonni}}, \bibnamefont{and}
  \bibinfo{author}{\bibfnamefont{P.}~\bibnamefont{Calabrese}},
  \bibinfo{journal}{J. Stat. Mech} \bibinfo{eid}{P08005}
  (\bibinfo{year}{2015}).

\bibitem[{\citenamefont{Coser et~al.}()\citenamefont{Coser, Tonni, and
  Calabrese}}]{CTC15b}
\bibinfo{author}{\bibfnamefont{A.}~\bibnamefont{Coser}},
  \bibinfo{author}{\bibfnamefont{E.}~\bibnamefont{Tonni}}, \bibnamefont{and}
  \bibinfo{author}{\bibfnamefont{P.}~\bibnamefont{Calabrese}},
  \bibinfo{note}{arXiv:1508.00811}.

\bibitem[{\citenamefont{Plenio et~al.}(2005)\citenamefont{Plenio, Eisert,
  Dreissig, and Cramer}}]{PEDC05}
\bibinfo{author}{\bibfnamefont{M.~B.} \bibnamefont{Plenio}},
  \bibinfo{author}{\bibfnamefont{J.}~\bibnamefont{Eisert}},
  \bibinfo{author}{\bibfnamefont{J.}~\bibnamefont{Dreissig}}, \bibnamefont{and}
  \bibinfo{author}{\bibfnamefont{M.}~\bibnamefont{Cramer}},
  \bibinfo{journal}{Phys. Rev. Lett.} \textbf{\bibinfo{volume}{94}},
  \bibinfo{pages}{060503} (\bibinfo{year}{2005}).

\bibitem[{\citenamefont{Cramer et~al.}(2007)\citenamefont{Cramer, Eisert, and
  Plenio}}]{CEP07}
\bibinfo{author}{\bibfnamefont{M.}~\bibnamefont{Cramer}},
  \bibinfo{author}{\bibfnamefont{J.}~\bibnamefont{Eisert}}, \bibnamefont{and}
  \bibinfo{author}{\bibfnamefont{M.~B.} \bibnamefont{Plenio}},
  \bibinfo{journal}{Phys. Rev. Lett.} \textbf{\bibinfo{volume}{98}},
  \bibinfo{pages}{220603} (\bibinfo{year}{2007}).

\bibitem[{\citenamefont{Wolf}(2006)}]{Wolf06}
\bibinfo{author}{\bibfnamefont{M.~M.} \bibnamefont{Wolf}},
  \bibinfo{journal}{Phys. Rev. Lett.} \textbf{\bibinfo{volume}{96}},
  \bibinfo{pages}{010404} (\bibinfo{year}{2006}).

\bibitem[{\citenamefont{Gioev and Klich}(2006)}]{GK06}
\bibinfo{author}{\bibfnamefont{D.}~\bibnamefont{Gioev}} \bibnamefont{and}
  \bibinfo{author}{\bibfnamefont{I.}~\bibnamefont{Klich}},
  \bibinfo{journal}{Phys. Rev. Lett.} \textbf{\bibinfo{volume}{96}},
  \bibinfo{pages}{100503} (\bibinfo{year}{2006}).

\bibitem[{\citenamefont{Barthel et~al.}(2006)\citenamefont{Barthel,
  \relax{M-C.} Chung, and Schollw\"ock}}]{BCS06}
\bibinfo{author}{\bibfnamefont{T.}~\bibnamefont{Barthel}},
  \bibinfo{author}{\bibnamefont{\relax{M-C.} Chung}}, \bibnamefont{and}
  \bibinfo{author}{\bibfnamefont{U.}~\bibnamefont{Schollw\"ock}},
  \bibinfo{journal}{Phys. Rev. A} \textbf{\bibinfo{volume}{74}},
  \bibinfo{pages}{022329} (\bibinfo{year}{2006}).

\bibitem[{\citenamefont{Farkas and Zimbor{\'a}s}(2007)}]{FZ07}
\bibinfo{author}{\bibfnamefont{S.}~\bibnamefont{Farkas}} \bibnamefont{and}
  \bibinfo{author}{\bibfnamefont{Z.}~\bibnamefont{Zimbor{\'a}s}},
  \bibinfo{journal}{J. Math. Phys.} \textbf{\bibinfo{volume}{48}},
  \bibinfo{pages}{102110} (\bibinfo{year}{2007}).

\bibitem[{\citenamefont{Swingle}(2010)}]{Swingle10}
\bibinfo{author}{\bibfnamefont{B.}~\bibnamefont{Swingle}},
  \bibinfo{journal}{Phys. Rev. Lett.} \textbf{\bibinfo{volume}{105}},
  \bibinfo{pages}{050502} (\bibinfo{year}{2010}).

\bibitem[{\citenamefont{Peres}(1996)}]{Peres96}
\bibinfo{author}{\bibfnamefont{A.}~\bibnamefont{Peres}},
  \bibinfo{journal}{Phys. Rev. Lett.} \textbf{\bibinfo{volume}{77}},
  \bibinfo{pages}{1413} (\bibinfo{year}{1996}).

\bibitem[{\citenamefont{Horodecki et~al.}(1996)\citenamefont{Horodecki,
  Horodecki, and Horodecki}}]{H3}
\bibinfo{author}{\bibfnamefont{M.}~\bibnamefont{Horodecki}},
  \bibinfo{author}{\bibfnamefont{P.}~\bibnamefont{Horodecki}},
  \bibnamefont{and}
  \bibinfo{author}{\bibfnamefont{R.}~\bibnamefont{Horodecki}},
  \bibinfo{journal}{Phys. Lett. A} \textbf{\bibinfo{volume}{223}},
  \bibinfo{pages}{1} (\bibinfo{year}{1996}).

\bibitem[{\citenamefont{Ferraro et~al.}(2008)\citenamefont{Ferraro, Cavalcanti,
  Garc\'ia-Saez, and Ac\'in}}]{FCGA08}
\bibinfo{author}{\bibfnamefont{A.}~\bibnamefont{Ferraro}},
  \bibinfo{author}{\bibfnamefont{D.}~\bibnamefont{Cavalcanti}},
  \bibinfo{author}{\bibfnamefont{A.}~\bibnamefont{Garc\'ia-Saez}},
  \bibnamefont{and} \bibinfo{author}{\bibfnamefont{A.}~\bibnamefont{Ac\'in}},
  \bibinfo{journal}{Phys. Rev. Lett.} \textbf{\bibinfo{volume}{100}},
  \bibinfo{pages}{080502} (\bibinfo{year}{2008}).

\bibitem[{\citenamefont{Anders and Winter}(2008)}]{AW08}
\bibinfo{author}{\bibfnamefont{J.}~\bibnamefont{Anders}} \bibnamefont{and}
  \bibinfo{author}{\bibfnamefont{A.}~\bibnamefont{Winter}},
  \bibinfo{journal}{Quantum Inf. Comput.} \textbf{\bibinfo{volume}{8}},
  \bibinfo{pages}{0245} (\bibinfo{year}{2008}).

\bibitem[{\citenamefont{Anders}(2008)}]{Anders08}
\bibinfo{author}{\bibfnamefont{J.}~\bibnamefont{Anders}},
  \bibinfo{journal}{Phys. Rev. A} \textbf{\bibinfo{volume}{77}},
  \bibinfo{pages}{062102} (\bibinfo{year}{2008}).

\bibitem[{\citenamefont{Marcovitch et~al.}(2009)\citenamefont{Marcovitch,
  Retzker, Plenio, and Reznik}}]{MRPR09}
\bibinfo{author}{\bibfnamefont{S.}~\bibnamefont{Marcovitch}},
  \bibinfo{author}{\bibfnamefont{A.}~\bibnamefont{Retzker}},
  \bibinfo{author}{\bibfnamefont{M.~B.} \bibnamefont{Plenio}},
  \bibnamefont{and} \bibinfo{author}{\bibfnamefont{B.}~\bibnamefont{Reznik}},
  \bibinfo{journal}{Phys. Rev, A} \textbf{\bibinfo{volume}{80}},
  \bibinfo{pages}{012325} (\bibinfo{year}{2009}).

\bibitem[{\citenamefont{Eisler and Zimbor\'as}(2014)}]{EZ14}
\bibinfo{author}{\bibfnamefont{V.}~\bibnamefont{Eisler}} \bibnamefont{and}
  \bibinfo{author}{\bibfnamefont{Z.}~\bibnamefont{Zimbor\'as}},
  \bibinfo{journal}{New J. Phys.} \textbf{\bibinfo{volume}{16}},
  \bibinfo{pages}{123020} (\bibinfo{year}{2014}).

\bibitem[{\citenamefont{Calabrese et~al.}(2015)\citenamefont{Calabrese, Cardy,
  and Tonni}}]{CCT15}
\bibinfo{author}{\bibfnamefont{P.}~\bibnamefont{Calabrese}},
  \bibinfo{author}{\bibfnamefont{J.}~\bibnamefont{Cardy}}, \bibnamefont{and}
  \bibinfo{author}{\bibfnamefont{E.}~\bibnamefont{Tonni}}, \bibinfo{journal}{J.
  Phys. A} \textbf{\bibinfo{volume}{48}}, \bibinfo{eid}{015006}
  (\bibinfo{year}{2015}).

\bibitem[{\citenamefont{Coser et~al.}(2014{\natexlab{a}})\citenamefont{Coser,
  Tonni, and Calabrese}}]{CTC14}
\bibinfo{author}{\bibfnamefont{A.}~\bibnamefont{Coser}},
  \bibinfo{author}{\bibfnamefont{E.}~\bibnamefont{Tonni}}, \bibnamefont{and}
  \bibinfo{author}{\bibfnamefont{P.}~\bibnamefont{Calabrese}},
  \bibinfo{journal}{J. Stat. Mech} \bibinfo{eid}{P12017}
  (\bibinfo{year}{2014}{\natexlab{a}}).

\bibitem[{\citenamefont{Hoogeveen and Doyon}(2015)}]{HD15}
\bibinfo{author}{\bibfnamefont{M.}~\bibnamefont{Hoogeveen}} \bibnamefont{and}
  \bibinfo{author}{\bibfnamefont{B.}~\bibnamefont{Doyon}},
  \bibinfo{journal}{Nucl. Phys. B} \textbf{\bibinfo{volume}{898}},
  \bibinfo{pages}{78} (\bibinfo{year}{2015}).

\bibitem[{\citenamefont{Wen et~al.}(2015)\citenamefont{Wen, Chang, and
  Ryu}}]{WCR15}
\bibinfo{author}{\bibfnamefont{X.}~\bibnamefont{Wen}},
  \bibinfo{author}{\bibfnamefont{P.-Y.} \bibnamefont{Chang}}, \bibnamefont{and}
  \bibinfo{author}{\bibfnamefont{S.}~\bibnamefont{Ryu}},
  \bibinfo{journal}{Phys. Rev. B} \textbf{\bibinfo{volume}{92}},
  \bibinfo{pages}{075109} (\bibinfo{year}{2015}).

\bibitem[{\citenamefont{Blondeau-Fournier
  et~al.}()\citenamefont{Blondeau-Fournier, Castro-Alvaredo, and
  Doyon}}]{BCD15}
\bibinfo{author}{\bibfnamefont{O.}~\bibnamefont{Blondeau-Fournier}},
  \bibinfo{author}{\bibfnamefont{O.~A.} \bibnamefont{Castro-Alvaredo}},
  \bibnamefont{and} \bibinfo{author}{\bibfnamefont{B.}~\bibnamefont{Doyon}},
  \bibinfo{journal}{J. Phys. A: Math. Theor.} \textbf{\bibinfo{volume}{49}},
  \bibinfo{pages}{125401} (\bibinfo{year}{2016}).

\bibitem[{\citenamefont{Alba}(2013)}]{Alba13}
\bibinfo{author}{\bibfnamefont{V.}~\bibnamefont{Alba}}, \bibinfo{journal}{J.
  Stat. Mech.} \bibinfo{eid}{P05013} (\bibinfo{year}{2013}).

\bibitem[{\citenamefont{Coser et~al.}(2014{\natexlab{b}})\citenamefont{Coser,
  Tagliacozzo, and Tonni}}]{CTT14}
\bibinfo{author}{\bibfnamefont{A.}~\bibnamefont{Coser}},
  \bibinfo{author}{\bibfnamefont{L.}~\bibnamefont{Tagliacozzo}},
  \bibnamefont{and} \bibinfo{author}{\bibfnamefont{E.}~\bibnamefont{Tonni}},
  \bibinfo{journal}{J. Stat. Mech} \bibinfo{eid}{P01008}
  (\bibinfo{year}{2014}{\natexlab{b}}).

\bibitem[{\citenamefont{Ju et~al.}(2012)\citenamefont{Ju, Kallin, Fendley,
  Hastings, and Melko}}]{Melko12}
\bibinfo{author}{\bibfnamefont{H.}~\bibnamefont{Ju}},
  \bibinfo{author}{\bibfnamefont{A.~B.} \bibnamefont{Kallin}},
  \bibinfo{author}{\bibfnamefont{P.}~\bibnamefont{Fendley}},
  \bibinfo{author}{\bibfnamefont{M.~B.} \bibnamefont{Hastings}},
  \bibnamefont{and} \bibinfo{author}{\bibfnamefont{R.~G.} \bibnamefont{Melko}},
  \bibinfo{journal}{Phys. Rev. B} \textbf{\bibinfo{volume}{85}},
  \bibinfo{pages}{165121} (\bibinfo{year}{2012}).

\bibitem[{\citenamefont{Leschke et~al.}(2014)\citenamefont{Leschke, Sobolev,
  and Spitzer}}]{LSS14}
\bibinfo{author}{\bibfnamefont{H.}~\bibnamefont{Leschke}},
  \bibinfo{author}{\bibfnamefont{A.~V.} \bibnamefont{Sobolev}},
  \bibnamefont{and} \bibinfo{author}{\bibfnamefont{W.}~\bibnamefont{Spitzer}},
  \bibinfo{journal}{Phys. Rev. Lett.} \textbf{\bibinfo{volume}{112}},
  \bibinfo{pages}{160403} (\bibinfo{year}{2014}).

\bibitem[{\citenamefont{Ding et~al.}(2012)\citenamefont{Ding, Seidel, and
  Yang}}]{DSY12}
\bibinfo{author}{\bibfnamefont{W.}~\bibnamefont{Ding}},
  \bibinfo{author}{\bibfnamefont{A.}~\bibnamefont{Seidel}}, \bibnamefont{and}
  \bibinfo{author}{\bibfnamefont{K.}~\bibnamefont{Yang}},
  \bibinfo{journal}{Phys. Rev. X} \textbf{\bibinfo{volume}{2}},
  \bibinfo{pages}{011012} (\bibinfo{year}{2012}).

\bibitem[{\citenamefont{McMinis and Tubman}(2013)}]{MT13}
\bibinfo{author}{\bibfnamefont{J.}~\bibnamefont{McMinis}} \bibnamefont{and}
  \bibinfo{author}{\bibfnamefont{N.~M.} \bibnamefont{Tubman}},
  \bibinfo{journal}{Phys. Rev. B} \textbf{\bibinfo{volume}{87}},
  \bibinfo{pages}{081108(R)} (\bibinfo{year}{2013}).

\bibitem[{\citenamefont{Keating and Mezzadri}(2005)}]{KM05}
\bibinfo{author}{\bibfnamefont{J.~P.} \bibnamefont{Keating}} \bibnamefont{and}
  \bibinfo{author}{\bibfnamefont{F.}~\bibnamefont{Mezzadri}},
  \bibinfo{journal}{Phys. Rev. Lett.} \textbf{\bibinfo{volume}{94}},
  \bibinfo{pages}{050501} (\bibinfo{year}{2005}).

\bibitem[{\citenamefont{Eisler and Zimbor\'as}(2005)}]{EZ05}
\bibinfo{author}{\bibfnamefont{V.}~\bibnamefont{Eisler}} \bibnamefont{and}
  \bibinfo{author}{\bibfnamefont{Z.}~\bibnamefont{Zimbor\'as}},
  \bibinfo{journal}{Phys. Rev. A} \textbf{\bibinfo{volume}{71}},
  \bibinfo{pages}{042318} (\bibinfo{year}{2005}).

\bibitem[{\citenamefont{K\'ad\'ar and Zimbor\'as}(2010)}]{KZ10}
\bibinfo{author}{\bibfnamefont{Z.}~\bibnamefont{K\'ad\'ar}} \bibnamefont{and}
  \bibinfo{author}{\bibfnamefont{Z.}~\bibnamefont{Zimbor\'as}},
  \bibinfo{journal}{Phys. Rev. A} \textbf{\bibinfo{volume}{82}},
  \bibinfo{pages}{032334} (\bibinfo{year}{2010}).

\bibitem[{\citenamefont{Ares et~al.}(2014)\citenamefont{Ares, Esteve, Falceto,
  and S{\'a}nchez-Burillo}}]{AEFS14}
\bibinfo{author}{\bibfnamefont{F.}~\bibnamefont{Ares}},
  \bibinfo{author}{\bibfnamefont{J.~G.} \bibnamefont{Esteve}},
  \bibinfo{author}{\bibfnamefont{F.}~\bibnamefont{Falceto}}, \bibnamefont{and}
  \bibinfo{author}{\bibfnamefont{E.}~\bibnamefont{S{\'a}nchez-Burillo}},
  \bibinfo{journal}{J. Phys. A: Math. Theor.} \textbf{\bibinfo{volume}{47}},
  \bibinfo{pages}{245301} (\bibinfo{year}{2014}).

\bibitem[{\citenamefont{Swingle}(2012)}]{Swingle12}
\bibinfo{author}{\bibfnamefont{B.}~\bibnamefont{Swingle}},
  \bibinfo{journal}{Phys. Rev. B} \textbf{\bibinfo{volume}{86}},
  \bibinfo{pages}{035116} (\bibinfo{year}{2012}).

\bibitem[{\citenamefont{Chandran et~al.}()\citenamefont{Chandran, Laumann, and
  Sorkin}}]{Sorkin15}
\bibinfo{author}{\bibfnamefont{A.}~\bibnamefont{Chandran}},
  \bibinfo{author}{\bibfnamefont{C.}~\bibnamefont{Laumann}}, \bibnamefont{and}
  \bibinfo{author}{\bibfnamefont{R.~D.} \bibnamefont{Sorkin}},
  \bibinfo{note}{arXiv:1511.02996}.

\bibitem[{\citenamefont{Li et~al.}(2006)\citenamefont{Li, Ding, Yu, Roscilde,
  and Haas}}]{LDYRH06}
\bibinfo{author}{\bibfnamefont{W.}~\bibnamefont{Li}},
  \bibinfo{author}{\bibfnamefont{L.}~\bibnamefont{Ding}},
  \bibinfo{author}{\bibfnamefont{R.}~\bibnamefont{Yu}},
  \bibinfo{author}{\bibfnamefont{T.}~\bibnamefont{Roscilde}}, \bibnamefont{and}
  \bibinfo{author}{\bibfnamefont{S.}~\bibnamefont{Haas}},
  \bibinfo{journal}{Phys. Rev. B} \textbf{\bibinfo{volume}{74}},
  \bibinfo{pages}{073103} (\bibinfo{year}{2006}).

\bibitem[{\citenamefont{Calabrese
  et~al.}(2012{\natexlab{b}})\citenamefont{Calabrese, Mintchev, and
  Vicari}}]{CMV12}
\bibinfo{author}{\bibfnamefont{P.}~\bibnamefont{Calabrese}},
  \bibinfo{author}{\bibfnamefont{M.}~\bibnamefont{Mintchev}}, \bibnamefont{and}
  \bibinfo{author}{\bibfnamefont{E.}~\bibnamefont{Vicari}},
  \bibinfo{journal}{EPL} \textbf{\bibinfo{volume}{97}}, \bibinfo{pages}{20009}
  (\bibinfo{year}{2012}{\natexlab{b}}).

\bibitem[{\citenamefont{Rodney et~al.}(2013)\citenamefont{Rodney, Song,
  \relax{S-S.} Lee, \relax{Le Hur}, and S{\o}rensen}}]{RSLHS13}
\bibinfo{author}{\bibfnamefont{M.}~\bibnamefont{Rodney}},
  \bibinfo{author}{\bibfnamefont{H.~F.} \bibnamefont{Song}},
  \bibinfo{author}{\bibnamefont{\relax{S-S.} Lee}},
  \bibinfo{author}{\bibfnamefont{K.}~\bibnamefont{\relax{Le Hur}}},
  \bibnamefont{and} \bibinfo{author}{\bibfnamefont{E.~S.}
  \bibnamefont{S{\o}rensen}}, \bibinfo{journal}{Phys. Rev. B}
  \textbf{\bibinfo{volume}{87}}, \bibinfo{pages}{115132}
  (\bibinfo{year}{2013}).

\bibitem[{\citenamefont{\relax{H-H.} Lai
  et~al.}(2013)\citenamefont{\relax{H-H.} Lai, Yang, and Bonesteel}}]{LYB13}
\bibinfo{author}{\bibnamefont{\relax{H-H.} Lai}},
  \bibinfo{author}{\bibfnamefont{K.}~\bibnamefont{Yang}}, \bibnamefont{and}
  \bibinfo{author}{\bibfnamefont{N.~E.} \bibnamefont{Bonesteel}},
  \bibinfo{journal}{Phys. Rev. Lett.} \textbf{\bibinfo{volume}{111}},
  \bibinfo{pages}{210402} (\bibinfo{year}{2013}).

\bibitem[{\citenamefont{\relax{H-H.} Lai and Yang}()}]{LY15}
\bibinfo{author}{\bibnamefont{\relax{H-H.} Lai}} \bibnamefont{and}
  \bibinfo{author}{\bibfnamefont{K.}~\bibnamefont{Yang}},
  \bibinfo{note}{arXiv:1510.03428}.

\bibitem[{\citenamefont{Peschel}(2003)}]{Peschel03}
\bibinfo{author}{\bibfnamefont{I.}~\bibnamefont{Peschel}}, \bibinfo{journal}{J.
  Phys. A: Math. Gen.} \textbf{\bibinfo{volume}{36}}, \bibinfo{pages}{L205}
  (\bibinfo{year}{2003}).

\bibitem[{\citenamefont{Peschel and Eisler}(2009)}]{PE09}
\bibinfo{author}{\bibfnamefont{I.}~\bibnamefont{Peschel}} \bibnamefont{and}
  \bibinfo{author}{\bibfnamefont{V.}~\bibnamefont{Eisler}},
  \bibinfo{journal}{J. Phys. A: Math. Theor.} \textbf{\bibinfo{volume}{42}},
  \bibinfo{pages}{504003} (\bibinfo{year}{2009}).

\bibitem[{\citenamefont{Cardy and Calabrese}(2010)}]{CC10}
\bibinfo{author}{\bibfnamefont{J.}~\bibnamefont{Cardy}} \bibnamefont{and}
  \bibinfo{author}{\bibfnamefont{P.}~\bibnamefont{Calabrese}},
  \bibinfo{journal}{J. Stat. Mech.} \bibinfo{eid}{P04023}
  (\bibinfo{year}{2010}).

\bibitem[{\citenamefont{Jin and Korepin}(2004)}]{JK04}
\bibinfo{author}{\bibfnamefont{B.~Q.} \bibnamefont{Jin}} \bibnamefont{and}
  \bibinfo{author}{\bibfnamefont{V.~E.} \bibnamefont{Korepin}},
  \bibinfo{journal}{J. Stat. Phys.} \textbf{\bibinfo{volume}{116}},
  \bibinfo{pages}{79} (\bibinfo{year}{2004}).

\bibitem[{\citenamefont{Wichterich et~al.}(2009)\citenamefont{Wichterich,
  Molina-Vilaplana, and Bose}}]{WMB09}
\bibinfo{author}{\bibfnamefont{H.}~\bibnamefont{Wichterich}},
  \bibinfo{author}{\bibfnamefont{J.}~\bibnamefont{Molina-Vilaplana}},
  \bibnamefont{and} \bibinfo{author}{\bibfnamefont{S.}~\bibnamefont{Bose}},
  \bibinfo{journal}{Phys. Rev. A} \textbf{\bibinfo{volume}{80}},
  \bibinfo{pages}{010304(R)} (\bibinfo{year}{2009}).

\bibitem[{\citenamefont{Bayat et~al.}(2010)\citenamefont{Bayat, Sodano, and
  Bose}}]{Bayat10}
\bibinfo{author}{\bibfnamefont{A.}~\bibnamefont{Bayat}},
  \bibinfo{author}{\bibfnamefont{P.}~\bibnamefont{Sodano}}, \bibnamefont{and}
  \bibinfo{author}{\bibfnamefont{S.}~\bibnamefont{Bose}},
  \bibinfo{journal}{Phys. Rev. B} \textbf{\bibinfo{volume}{81}},
  \bibinfo{pages}{064429} (\bibinfo{year}{2010}).

\bibitem[{\citenamefont{Calabrese
  et~al.}(2013{\natexlab{b}})\citenamefont{Calabrese, Tagliacozzo, and
  Tonni}}]{CTT13}
\bibinfo{author}{\bibfnamefont{P.}~\bibnamefont{Calabrese}},
  \bibinfo{author}{\bibfnamefont{L.}~\bibnamefont{Tagliacozzo}},
  \bibnamefont{and} \bibinfo{author}{\bibfnamefont{E.}~\bibnamefont{Tonni}},
  \bibinfo{journal}{J. Stat. Mech.} \bibinfo{eid}{P05002}
  (\bibinfo{year}{2013}{\natexlab{b}}).

\bibitem[{\citenamefont{Chung et~al.}(2014)\citenamefont{Chung, Alba, Bonnes,
  Chen, and L\"auchli}}]{Chung14}
\bibinfo{author}{\bibfnamefont{C.-M.} \bibnamefont{Chung}},
  \bibinfo{author}{\bibfnamefont{V.}~\bibnamefont{Alba}},
  \bibinfo{author}{\bibfnamefont{L.}~\bibnamefont{Bonnes}},
  \bibinfo{author}{\bibfnamefont{P.}~\bibnamefont{Chen}}, \bibnamefont{and}
  \bibinfo{author}{\bibfnamefont{A.~M.} \bibnamefont{L\"auchli}},
  \bibinfo{journal}{Phys. Rev. B} \textbf{\bibinfo{volume}{90}},
  \bibinfo{pages}{064401} (\bibinfo{year}{2014}).

\bibitem[{\citenamefont{Sherman et~al.}()\citenamefont{Sherman, Devakul,
  Hastings, and Singh}}]{SDHS15}
\bibinfo{author}{\bibfnamefont{N.~E.} \bibnamefont{Sherman}},
  \bibinfo{author}{\bibfnamefont{T.}~\bibnamefont{Devakul}},
  \bibinfo{author}{\bibfnamefont{M.~B.} \bibnamefont{Hastings}},
  \bibnamefont{and} \bibinfo{author}{\bibfnamefont{R.~R.~P.}
  \bibnamefont{Singh}}, \bibinfo{note}{arXiv:1510.08005}.

\bibitem[{\citenamefont{Lee and Vidal}(2013)}]{LV13}
\bibinfo{author}{\bibfnamefont{Y.~A.} \bibnamefont{Lee}} \bibnamefont{and}
  \bibinfo{author}{\bibfnamefont{G.}~\bibnamefont{Vidal}},
  \bibinfo{journal}{Phys. Rev. A} \textbf{\bibinfo{volume}{88}},
  \bibinfo{pages}{042318} (\bibinfo{year}{2013}).

\bibitem[{\citenamefont{Castelnovo}(2013)}]{Castel13}
\bibinfo{author}{\bibfnamefont{C.}~\bibnamefont{Castelnovo}},
  \bibinfo{journal}{Phys. Rev. A} \textbf{\bibinfo{volume}{88}},
  \bibinfo{pages}{042319} (\bibinfo{year}{2013}).

\bibitem[{\citenamefont{Klich}()}]{Klich02}
\bibinfo{author}{\bibfnamefont{I.}~\bibnamefont{Klich}},
  \bibinfo{note}{arXiv:cond-mat/0209642}.

\bibitem[{\citenamefont{Fagotti and Calabrese}(2010)}]{FC10}
\bibinfo{author}{\bibfnamefont{M.}~\bibnamefont{Fagotti}} \bibnamefont{and}
  \bibinfo{author}{\bibfnamefont{P.}~\bibnamefont{Calabrese}},
  \bibinfo{journal}{J. Stat. Mech.} \bibinfo{eid}{P04016}
  (\bibinfo{year}{2010}).

\bibitem[{\citenamefont{Klich}(2014)}]{Klich14}
\bibinfo{author}{\bibfnamefont{I.}~\bibnamefont{Klich}}, \bibinfo{journal}{J.
  Stat. Mech.} \bibinfo{eid}{P11006} (\bibinfo{year}{2014}).

\end{thebibliography}

\end{document}